\documentclass[11pt,a4paper]{article}
\usepackage{jheppub}
\usepackage[T1]{fontenc}
\usepackage{fix-cm}
\usepackage{lmodern}
\usepackage{amsmath,amssymb,mathtools,mathrsfs,empheq}
\usepackage{physics}
\usepackage{bbm}
\usepackage{bm}
\usepackage{hhline}

\newcommand\nn{\nonumber}

\makeatletter
\newcommand*{\rom}[1]{\expandafter\@slowromancap\romannumeral #1@}
\makeatother

\def\ri{{\rm i}}

\usepackage{tikz}

\usepackage{xcolor}

\newcommand\fft[2]{\frac{#1}{#2}}

\def\SO{{\rm SO}}

\def\SU{{\rm SU}}
\def\U{{\rm U}}

\usepackage{tensor}

\newcommand\mO{\mathcal{O}}
\newcommand\mA{\mathcal{A}}
\newcommand\tV{\widetilde{V}}

\newcommand\mV{\mathcal{V}}
\newcommand\mW{\mathcal{W}}
\newcommand\tc{\tilde{c}}
\newcommand\mY{\mathcal{Y}}
\newcommand\mS{\mathcal{S}}

\title{AdS$_7$ Black Holes from Rotating M5-branes}
\author[a]{Nikolay Bobev,}
\author[a]{Marina David,}
\author[a]{Junho Hong,}
\author[b]{and Rishi Mouland}

\affiliation[a]{Instituut voor Theoretische Fysica, KU Leuven, \\
	Celestijnenlaan 200D, B-3001 Leuven, Belgium}

\affiliation[b]{DAMTP, Centre for Mathematical Sciences, \\
University of Cambridge, Wilberforce Road
Cambridge CB3 0WA, UK}

\emailAdd{nikolay.bobev@kuleuven.be}
\emailAdd{junho.hong@kuleuven.be}
\emailAdd{marina.david@kuleuven.be}
\emailAdd{r.mouland@damtp.cam.ac.uk}

\abstract{We construct a general asymptotically locally AdS$_7$ stationary black hole solution of 7d maximal gauged supergravity with three angular momenta and two electric charges. When embedded in 11d supergravity the black hole  describes the backreaction of $N$ coincident rotating M5-branes. We study the thermodynamic properties of the black hole and present explicit expressions for its entropy, energy, electric charges, and angular momenta. We show that in the supersymmetric limit of the solution its entropy and on-shell action precisely agree with the result for the path integral of the holographically dual 6d $\mathcal{N}=(2,0)$ SCFT on $S^1\times S^5$ to leading order in the large $N$ limit.}

\begin{document}
	
\maketitle 

	
\section{Introduction}
\label{sec:intro}

Understanding the low-energy quantum dynamics of a system of $N$ coincident M5-branes in M-theory is an important and notoriously difficult open problem. This is especially true if the 6d $\mathcal{N}=(2,0)$ quantum field theory on the worldvolume of the M5-branes is at finite temperature and chemical potentials. The AdS/CFT correspondence offers a unique and calculationally powerful vantage point to shed light on this problem at large $N$. In this limit the dynamics of the system is expected to be described by an asymptotically AdS$_7$ black hole solution of 11d supergravity with non-vanishing angular momenta and electric charges. Since the 6d $\mathcal{N}=(2,0)$ theory is invariant under the $\SO(2,6)$ conformal and $\SO(5)$ R-symmetry it is expected that the most general black hole solution of this type is characterized by 6 independent conserved charges - the mass $M$, three angular momenta $J_{1,2,3}$, and two electric charges $Q_{1,2}$. The main goal of this work is to construct this black hole solution explicitly and study its properties in detail. 

The problem of constructing AdS$_7$ black hole solutions sourced by M5-branes is not new and has been studied in the literature. The first example of such a 7d black hole solution with three equal angular momenta and two different electric charges was presented in \cite{Chong:2004dy}. A different solution with two equal electric charges and three different angular momenta was derived in \cite{Chow:2007ts}. Finally, a solution with two vanishing angular momenta and two different electric charges was given in \cite{Wu:2011gp} and \cite{Chow:2011fh}. The black hole solution we find here generalizes these results and has arbitrary values for the mass, three angular momenta, and two electric charges. In the appropriate limits of these charges our results agree with the previous literature. 

Equipped with this novel explicit solution we proceed to study its properties and derive explicit expressions for its mass, three angular momenta, two electric charges as well as its temperature and entropy. We then use these results to demonstrate that the first law of black hole thermodynamics is obeyed. Via the AdS/CFT correspondence the thermal properties of the black hole inform the finite temperature physics of the 6d $\mathcal{N}=(2,0)$ SCFT at large $N$ and provide explicit predictions for its thermal observables. It is challenging at present to confirm these predictions by using independent QFT methods. The black hole solutions however possess a smooth supersymmetric limit where we can make contact with available results for the large $N$ physics of the 6d $\mathcal{N}=(2,0)$ theory.  

To understand this supersymmetric limit better it is useful to study the six-parameter family of black hole solutions in Euclidean signature. The utility of this Euclidean approach to the supersymmetric limit of AdS black holes was emphasized recently in similar contexts in \cite{Cassani:2019mms,Cabo-Bizet:2018ehj,Bobev:2019zmz,Bobev:2020pjk,Kantor:2019lfo}. After imposing a particular relation between the six parameters specifying the Euclidean background, we find a five-parameter family of solutions that enjoys a linear relation between the energy, three angular momenta, and two electric charges which is characteristic of supersymmetry. An important feature of this family of solutions is that they cap off smoothly in the IR region and therefore do not suffer from IR divergences and ambiguities in the evaluation of their thermodynamic charges and on-shell action. Indeed, using the first law of black hole thermodynamics, also known as the quantum statistical relation, on this family of supersymmetric solutions we find a simple expression for their on-shell action. Employing the standard holographic dictionary for asymptotically AdS$_7$ solutions sourced by M5-branes we show that this result for the regularized on-shell action precisely agrees with the results for the large $N$ limit of the superconformal index \cite{Bhattacharya:2008zy} of the dual 6d $\mathcal{N}=(2,0)$ $A_N$ SCFT \cite{Bobev:2015kza,Hosseini:2018dob,Choi:2018hmj,Nahmgoong:2019hko,Ohmori:2020wpk}. This constitutes a non-trivial precision test of holography and generalizes similar holographic calculations for AdS black holes in lower dimensions. 

The five-parameter family of supersymmetric Euclidean solutions does not admit a good analytic continuation to smooth, regular, and causal supersymmetric black holes in AdS$_7$. We find that in order to have a well-defined Lorentzian black hole solution with real and positive entropy we must impose an additional non-linear relation between the charges and angular momenta of the supergravity solution. After imposing this relation we find a four-parameter family of BPS black holes with positive entropy which we compute explicitly. The result precisely agrees with an appropriate Legendre transformation of the large $N$ limit of the superconformal index of the 6d $\mathcal{N}=(2,0)$ $A_N$ SCFT. This calculation can be viewed as a microscopic derivation of the semi-classical black hole entropy from the large $N$ limit of the dual SCFT. The non-linear relation between the black hole charges and angular momenta is a familiar predicament from the study of supersymmetric electrically charged rotating black holes in AdS$_4$ and AdS$_5$ and remains mysterious from the perspective of the dual SCFT.

We continue in the next section with a short discussion on the 7d gauged supergravity theory we study and how it arises as a consistent truncation of 11d supergravity. In Section~\ref{sec:BHsol} we present the black hole solution of interest and in Section~\ref{sec:TDOnshell} we study its thermodynamic properties and discuss how to take the supersymmetric limit in both Lorentzian and Euclidean signature. In Section~\ref{sec:holo} we show that our supergravity results are in precise agreement with the expectations based on the superconformal index of the 6d $\mathcal{N}=(2,0)$ SCFT on $S^1\times S^5$ to leading order in the large $N$ limit. In Section~\ref{sec:discussion} we discuss some open problems and directions for future work. Several technical details, as well as a discussion of various limits of our solution, are discussed in the appendices.


\section{7d gauged supergravity}
\label{sec:7dsugra}

The gravitational solution we are after is expected to solve the equations of motion of 7d maximal gauged supergravity. This theory arises as a consistent truncation of 11d supergravity on $S^4$ as shown in \cite{Nastase:1999cb,Nastase:1999kf}.\footnote{The theory can also arise from a consistent truncation on $\mathbb{RP}^4$ which is relevant for the holographic description of the 6d $D_N$ series of $\mathcal{N}=(2,0)$ SCFTs.} The solution of interest preserves a $\U(1) \times \U(1)$ subgroup of the $\SO(5)$ symmetry of the gauged supergravity and therefore can be found in a further $\U(1) \times \U(1)$ invariant consistent truncation that was worked out in \cite{Liu:1999ai}. The bosonic fields in this consistent truncation are the metric, two Abelian gauge fields, two real scalars and a single 3-form potential with a ``self-dual'' 4-form flux.

The bosonic action of the $\U(1) \times \U(1)$ invariant consistent truncation in Lorentzian signature reads\footnote{We use the same Lagrangian and equations of motion as the ones employed in \cite{Chow:2007ts}. Our conventions for differential forms are summarized in Appendix~\ref{App:convention}.}
\begin{align}\label{eq:S7d}
	S&=\fft{1}{16\pi G_N}\int\bigg[R \star 1 +2g^2\bigg(8X_1X_2+\fft{4(X_1+X_2)}{X_1^2X_2^2}-\fft{1}{X_1^4X_2^4}\bigg)\star 1 -\fft12\sum_{I=1}^2d\varphi_I\wedge \star d\varphi_I\nn\\
	&\kern6em~-\fft12\sum_{I=1}^2\fft{1}{X_I^2}F^I_{(2)}\wedge \star F^{I}_{(2)}-\fft{1}{2}X_1^2X_2^2F_{(4)}\wedge \star F_{(4)}\nn\\
	&\kern6em~+gF_{(4)}\wedge A_{(3)}+F^1_{(2)}\wedge F^2_{(2)}\wedge A_{(3)}\bigg]\,,
\end{align}
where we have defined
\begin{equation}
X_1 = e^{-\frac{1}{\sqrt{10}}\varphi_1-\frac{1}{\sqrt{2}}\varphi_2}\,, \quad X_2 = e^{-\frac{1}{\sqrt{10}}\varphi_1+\frac{1}{\sqrt{2}}\varphi_2}\,, \quad F_{(2)}^{I} = dA_{(1)}^{I}\,, \quad F_{(4)} = dA_{(3)}\,.
\end{equation}
It is straightforward to derive the bosonic equations of motion for this action. The Einstein equations read
\begin{equation}
\begin{split}
	0&=R_{\mu\nu}-\fft12g_{\mu\nu}\bigg[R+2g^2\bigg(8X_1X_2+\fft{4(X_1+X_2)}{X_1^2X_2^2}-\fft{1}{X_1^4X_2^4}\bigg)\bigg]\\
	&\quad-\sum_{I=1}^2\bigg[\fft12\partial_\mu\varphi_I\partial_\nu\varphi_I-\fft14g_{\mu\nu}\partial^\rho\varphi_I\partial_\rho\varphi_I\bigg]-\sum_{I=1}^2\fft{1}{X_I^2}\bigg[\fft12F^I_{\mu\rho}F^I_{\nu}{}^\rho-\fft18g_{\mu\nu}F^I_{\rho\sigma}F^{I\,\rho\sigma}\bigg]\\
	&\quad-X_1^2X_2^2\bigg[\fft{1}{12}F_{\mu\rho\sigma\lambda}F_{\nu}{}^{\rho\sigma\lambda}-\fft{1}{96}g_{\mu\nu}F_{\rho\sigma\lambda\delta}F^{\rho\sigma\lambda\delta}\bigg]\,.
\end{split}\label{eom:Einstein}
\end{equation}
The scalar equations of motion are given by
\begin{equation}
\begin{split}
	\Box\varphi_1&=\fft{1}{2\sqrt{10}}\sum_{I=1}^2\fft{1}{X_I^2}F^I_{\mu\nu}F^{I\,\mu\nu}-\fft{1}{12\sqrt{10}}X_1^2X_2^2F_{\mu\nu\rho\sigma}F^{\mu\nu\rho\sigma}\\
	&\quad+\fft{8g^2}{\sqrt{10}}\bigg(4X_1X_2-\fft{3(X_1+X_2)}{X_1^2X_2^2}+\fft{2}{X_1^4X_2^4}\bigg)\,,\\
	\Box\varphi_2&=\fft{1}{2\sqrt{2}}\bigg[\fft{1}{X_1^2}F^1_{\mu\nu}F^{1\,\mu\nu}-\fft{1}{X_2^2}F^2_{\mu\nu}F^{2\,\mu\nu}\bigg]+4\sqrt{2}g^2\fft{X_1-X_2}{X_1^2X_2^2}\,,
\end{split}\label{eom:scalar}
\end{equation}
where we have defined $\Box\equiv\nabla^\mu\nabla_\mu$. The vector and 3-form equations of motion take the form
\begin{subequations}
\begin{align}
	d(X_1^{-2} \star F^1_{(2)})&=F^2_{(2)}\wedge F_{(4)}\,,\label{eom:vector1}\\
	d(X_2^{-2} \star F^2_{(2)})&=F^1_{(2)}\wedge F_{(4)}\,,\label{eom:vector2}\\
	d(X_1^2X_2^2 \star F_{(4)})&=2gF_{(4)}+F^1_{(2)}\wedge F^2_{(2)}\,.\label{eom:3form}
\end{align}
\end{subequations}
In addition to the equations of motion, we must impose a ``self-duality'' equation on $A_{(3)}$, see \cite{Pilch:1984xy}, that reads
\begin{equation}
	X_1^2X_2^2 \star F_{(4)}=2gA_{(3)}+\fft12\left(A^1_{(1)}\wedge F^2_{(2)}+A^2_{(1)}\wedge F^1_{(2)}\right)-dA_{(2)}\,,\label{self-dual}
\end{equation}
written in terms of a 2-form potential $A_{(2)}$. Note that the existence of the 2-form potential $A_{(2)}$ satisfying the self-duality equation (\ref{self-dual}) can be shown locally by applying the Poincar\'{e} lemma to the 3-form equation of motion (\ref{eom:3form}), and therefore the self-duality equation does not impose any extra condition on a local solution to the equations of motion.

This supergravity truncation admits a further interesting truncation which is obtained by setting $X_1=X_2$ and $A^1_{(1)}=A^1_{(2)}$. As explained carefully in \cite{Lu:1999bc,Nastase:2000tu} this model is equivalent to the $\U(1)$ invariant truncation of the minimal 7d $\SU(2)$ gauged supergravity theory of \cite{Townsend:1983kk}. We also note that the gravitational action in \eqref{eq:S7d} is a consistent truncation of 7d maximal $\SO(5)$ gauged supergravity which is not by itself a supergravity theory. To determine whether a given solution of the equations of motion presented above preserves supersymmetry we have to study the supersymmetry variations of the full 7d maximal $\SO(5)$ gauged supergravity which can be found in \cite{Pernici:1984xx,Liu:1999ai,Gauntlett:2000ng}.

\section{The black hole solution}
\label{sec:BHsol}

After some trial and error and using, as inspiration, the existing 7d black hole solutions in the literature \cite{Chong:2004dy,Chow:2007ts,Wu:2011gp,Chow:2011fh} we managed to find a general solution of the equations of motion presented in Section~\ref{sec:7dsugra} that has one mass, three rotation and two charge parameters. To present the solution we use the same $(t,r,y,z,\phi_1,\phi_2,\phi_3)$ coordinates employed in \cite{Chow:2007ts}. 

The metric reads

\begin{equation}
	\begin{split}
		ds^2&= (H_1H_2)^{\fft15}\left(-\fft{(1+g^2r^2)(1-g^2y^2)(1-g^2z^2)}{\Xi_1\Xi_2\Xi_3}dt^2+\fft{(r^2+y^2)(r^2+z^2)}{U}dr^2 \right. \\
		& \left. \quad+\fft{(r^2+y^2)(y^2-z^2)y^2}{(1-g^2y^2)(a_1^2-y^2)(a_2^2-y^2)(a_3^2-y^2)}dy^2 \right. \\
		& \left. \quad+\fft{(r^2+z^2)(z^2-y^2)z^2}{(1-g^2z^2)(a_1^2-z^2)(a_2^2-z^2)(a_3^2-z^2)}dz^2 \right. \\
		& \left. \quad+\fft{(a_1^2+r^2)(a_1^2-y^2)(a_1^2-z^2)}{\Xi_1(a_1^2-a_2^2)(a_1^2-a_3^2)}d\phi_1^2+\fft{(a_2^2+r^2)(a_2^2-y^2)(a_2^2-z^2)}{\Xi_2(a_2^2-a_1^2)(a_2^2-a_3^2)}d\phi_2^2 \right. \\
		&\left. \quad+\fft{(a_3^2+r^2)(a_3^2-y^2)(a_3^2-z^2)}{\Xi_3(a_3^2-a_1^2)(a_3^2-a_2^2)}d\phi_3^2+\fft{1-\fft{1}{H_1}}{1-(s_2/s_1)^2}K_1^2+\fft{1-\fft{1}{H_2}}{1-(s_1/s_2)^2}K_2^2 \right) \,, 
	\end{split}\label{ansatz:metric}
\end{equation}
%
%
where we have defined ($i,j,k\in\{1,2,3\}$ are all different)
\begin{subequations}
\begin{align}
		s_I&=\sinh\delta_I\,, \quad c_I=\cosh\delta_I\,, \\
		\Xi_i &= 1-a_i^2g^2\,, \\
		H_I(r,y,z)&=1+\fft{2ms_I^2}{(r^2+y^2)(r^2+z^2)}\,,\\
		U(r)&=\fft{(1+g^2r^2)\textstyle\prod_{i=1}^{3}(r^2+a_i^2)}{r^2}-2m+mg^2(s_1^2+s_2^2)(2r^2+  {\textstyle \sum_{i=1}^{3}}a_{i}^2) +\fft{4m^2g^2s_1^2s_2^2}{r^2} \nonumber \\
		&\quad-\fft{2mg(s_1^2+s_2^2)a_1a_2a_3}{r^2}+
		\fft{2mg(c_1-c_2)^2}{r^2}(a_1+a_2a_3g)(a_2+a_3a_1g)(a_3+a_1a_2g) \nonumber \\
		&\quad-\fft12mg^2(c_1-c_2)^2\Big(-2(a_1^2+a_2^2+a_3^2)-8a_1a_2a_3g \nonumber \\
		&\quad -(a_1+a_2+a_3)(a_2+a_3-a_1)(a_3+a_1-a_2)(a_1+a_2-a_3)g^2\Big)\,,\\
		K_1&=\fft{c_1+c_2}{2s_1}\mathcal{A}[y^2,z^2,0]+\fft{c_1-c_2}{2s_1}\mY\,,\\
		K_2&=\fft{c_1+c_2}{2s_2}\mathcal{A}[y^2,z^2,0]-\fft{c_1-c_2}{2s_2}\mY\,,\\
		\mathcal{A}[y^2,z^2,0]&=\fft{(1-g^2y^2)(1-g^2z^2)}{\Xi_1\Xi_2\Xi_3}dt-\sum_{i=1}^3\fft{a_i(a_i^2-y^2)(a_i^2-z^2)}{\Xi_i(a_i^2-a_j^2)(a_i^2-a_k^2)}d\phi_i\,,\\
		\mY&=\fft{(1-g^2y^2)(1-g^2z^2)(1-(a_1^2+a_2^2+a_3^2)g^2-2a_1a_2a_3g^3)}{\Xi_1\Xi_2\Xi_3}dt \nonumber\\
		&\quad+\sum_{i=1}^3\fft{a_i(a_i^2-y^2)(a_i^2-z^2)(1-(a_i^2-a_j^2-a_k^2)g^2+\fft{2a_ja_k}{a_i}g)}{\Xi_i(a_i^2-a_j^2)(a_i^2-a_k^2)}d\phi_i\,.
	\label{ansatz:functions}
\end{align}
\end{subequations}
Here $(m,a_1,a_2,a_3, \delta_1,\delta_2)$ are the six parameters that specify the solution.
The two Abelian gauge fields can be written as 1-forms given in terms of $K_{1,2}$ as
\begin{equation}\label{eq:A1Idef}
	A^I_{(1)}=\bigg(1-\fft{1}{H_I}\bigg)K_I + \alpha_I dt\,.
\end{equation}
The importance of introducing the constant pure gauge parameters $\alpha_I$ is emphasized later. The 3-form is given by
\begin{equation}
	\begin{split}
		A_{(3)}&=2ms_1s_2g^4a_1a_2a_3\bigg[\mA[y^2,z^2,0]-\mA[y^2,z^2,g^{-2}]\bigg]\\
		&\hspace{3mm}\wedge\bigg[\fft{dz\wedge\left(\mA[y^2,0,0]-\mA[y^2,0,g^{-2}]\right)}{(r^2+y^2)z}+\fft{dy\wedge\left(\mA[z^2,0,0]-\mA[z^2,0,g^{-2}]\right)}{(r^2+z^2)y}\bigg]\\
		&\hspace{3mm}+2ms_1s_2g^3\mA[y^2,z^2,0]\\
		&\hspace{3mm}\wedge\left[\fft{z \, dz\wedge\left(\mA[y^2,0,0]-\mA[y^2,0,g^{-2}]\right)}{(r^2+y^2)}+\fft{y\, dy\wedge\left(\mA[z^2,0,0]-\mA[z^2,0,g^{-2}]\right)}{(r^2+z^2)}\right]\,,\label{ansatz:3form}
	\end{split}
\end{equation}
%
%
where we have defined
\begin{equation}
	\mA[v_1,v_2,v_3]=\Bigg[\prod_{i=1}^3\fft{1-g^2v_i}{\Xi_i}\Bigg]dt-\sum_{i=1}^3\Bigg[\fft{(a_i^2-v_1)(a_i^2-v_2)(a_i^2-v_3)}{a_i\Xi_i\prod_{j=1\,(\neq i)}^3(a_i^2-a_j^2)}\Bigg]d\phi_i\,.
\end{equation}
Finally, the two scalars are given by
\begin{equation}
	X_I=\fft{(H_1H_2)^\fft25}{H_I}\,.\label{ansatz:scalar}
\end{equation}
These unwieldy expressions fully specify the supergravity solution. To check explicitly the self-duality constraint in \eqref{self-dual} we can use the following explicit expression for the 2-form potential
\begin{equation}
	\begin{split}
		A_{(2)}&=\bigg(\fft{1}{H_1}+\fft{1}{H_2}\bigg)\fft{ms_1s_2(a_1+a_2a_3g)}{(r^2+y^2)(r^2+z^2)}\bigg(\fft{(1-g^2y^2)(1-g^2z^2)\mu_1^2}{\Xi_1(1-a_2^2g^2)(1-a_3^2g^2)}dt\wedge d\phi_1\\
		&\kern16em+\fft{g(a_3^2-a_2^2)\mu_2^2\mu_3^2}{\Xi_2\Xi_3}d\phi_2\wedge d\phi_3\bigg)\\
		&\quad+\text{(cyclic-permutations)}\\
		&\quad - \frac{1}{2}dt\wedge \left(\alpha_1 A^2_{(1)} +\alpha_2 A^1_{(1)}\right)\,,\label{ansatz:2form}
	\end{split}
\end{equation}
where we have defined
\begin{equation}\label{def:mu}
	\mu_i^2=\fft{(a_i^2-y^2)(a_i^2-z^2)}{\prod_{j=1\,(\neq i)}^3(a_i^2-a_j^2)}\,,
\end{equation}
and recall the $\alpha_I$ are as-of-yet undetermined constants appearing in (\ref{eq:A1Idef}). Cyclic permutations signifies cycling over the rotation parameters $a_{i}$ and $\phi_{i}$ coordinates for $i=1,2,3$. 

We have checked explicitly that the expressions for all fields above satisfy the equations of motion \eqref{eom:Einstein}, \eqref{eom:scalar}, \eqref{eom:vector1}, \eqref{eom:vector2} and \eqref{eom:3form}, as summarized in Section~\ref{sec:7dsugra}.

Special cases of this solution --- with various equal parameters as well as vanishing parameters --- were previously found in the literature in \cite{Chong:2004dy,Chow:2007ts,Wu:2011gp,Chow:2011fh}. In Appendix~\ref{App:limits} we discuss in some detail these various specializations of our solution and provide a map between our parametrization of the solution and the ones used in the previous literature.

In certain limits of the parameters the solution presented above preserves some amount of supersymmetry. This is discussed in more detail in Section~\ref{sec:analysis:susy} where we also emphasize some subtleties on how to take the supersymmetric limits both in Euclidean and Lorentzian signatures.  

We note that the 7d background presented above can be uplifted to a solution of 11d supergravity by using the explicit uplift formulae in \cite{Nastase:1999cb,Nastase:1999kf,Cvetic:1999xp}. This 11d solution can be interpreted as the backreaction of a set of coincident rotating M5-branes. This perspective is important in the holographic discussion in Section~\ref{sec:holo}.

Finally, let us present coordinates that make manifest the AdS$_7$ asymptotics. We transform from $(t,r,y,z,\phi_i)$ to $(t,\tilde{r},\tilde{\mu}_i,\phi_i)$, where $\tilde{\mu}_i\ge 0$ with $\tilde{\mu}_1^2 + \tilde{\mu}_2^2 + \tilde{\mu}_3^2 =1$. These new coordinates are defined implicitly through the relations
\begin{align}\label{eq: spheroidal coord change}
  (1-g^2 a_i^2)\tilde{r}^2 \tilde{\mu}_i^2 = (r^2+a_i^2)\mu_i^2\,,
\end{align} 
with $\mu_i$ defined in terms of $(y,z)$ as in (\ref{def:mu}). Then at $\tilde{r}\to\infty$, the solution goes to AdS$_7$ in global coordinates, that is
\begin{align}
  ds^2 \sim \frac{d\tilde{r}^2}{g^2 \tilde{r}^2} + \tilde{r}^2 \left(- g^2 dt^2 + \sum_i \left(d\tilde{\mu}_i^2 + \tilde{\mu}_i^2 d\phi_i^2\right) \right) + \dots \,. 
\end{align}
\section{Black hole thermodynamics and on-shell action}
\label{sec:TDOnshell}

After presenting the explicit supergravity solution we are ready to discuss some of its physical properties. We present the solution in Lorentzian signature but it is useful to also analytically continue to Euclidean signature by taking $t \to - {\rm i} \tau$. We are mainly interested in computing the conserved charges and thermodynamic quantities of the supergravity solution above. To this end we need to specify the ranges of the 7d coordinates we use. The angles $\phi_{1,2,3}$ take values in the domain $[0,2\pi)$. The range of the coordinates $y$ and $z$ is finite and is determined by the magnitude of the parameters $a_{1,2,3}$. For simplicity we pick non-negative rotation parameters $a_i$'s\footnote{In general the rotation parameters $a_i$'s may take negative values. In that case thermodynamic quantities we have evaluated in this section are supposed to be slightly modified by taking absolute signs for some of the negative rotation parameters in the expressions carefully.} and take the following ranges
\begin{equation}\label{eq:yandzrange}
	0\leq a_1\leq z=\sqrt{a_1^2\cos^2\theta_2+a_2^2\sin^2\theta_2} \leq a_2\leq y=\sqrt{a_2^2\cos^2\theta_1+a_3^2\sin^2\theta_1} \leq a_3\leq g^{-1}\,,
\end{equation}
where we have defined the angular variables $\theta_{1,2}$ which take values in the domain $[0,\pi/2]$. The coordinates $\{y,z,\phi_1,\phi_2,\phi_3\}$ therefore span a topological $S^5$. When the supergravity solution is viewed as a black hole in Lorentzian signature the outer horizon is determined by the largest positive root of the function $U(r)$ in (\ref{ansatz:metric}), i.e. 
\begin{equation}
	r=r_+\quad\text{as the largest positive root of}\quad U(r_+)=0\,,
\end{equation}
and we take the range of $r$ to be from $r_+$ to the asymptotically locally AdS$_7$ boundary at $r \to \infty$. Finding a closed form expression for $r_+$ as a function of the 6 parameters $\{m,a_{1,2,3,},\delta_{1,2}\}$ determining the black hole solution is indeed possible but in practice, not advantageous to work with since the polynomial equation is of high degree. Therefore in many of the expressions below we keep $r_{+}$ as an implicit function of $\{m,a_{1,2,3,},\delta_{1,2}\}$. In Lorentzian signature the coordinate $t$ is infinite in range, while in Euclidean $\tau$ takes values in the range $[0,\beta]$. Therefore in Euclidean signature the asymptotic boundary of the solution has $S^1\times S^5$ topology.

\subsection{Black hole thermodynamics}\label{sec:analysis:THD}

To calculate the entropy of the black hole solution we simply need to calculate the area of the horizon which reads
\begin{equation}
\begin{split}
	S&=\fft{1}{4G_N}\int_{S_5}d^5x\,\sqrt{h}\,\Big|_{r=r_+}=\fft{1}{4G_N}\int_{a_1}^{a_2}dz\int_{a_2}^{a_3}dy\int_0^{2\pi}d\phi_1d\phi_2d\phi_3\,\sqrt{h}\,\Big|_{r=r_+}\\
	&=\fft{\pi^3}{4G_N\Xi_1\Xi_2\Xi_3}\fft{\sqrt{\mS(r_+)}}{r_+}\,.
\end{split}\label{eq:S}
\end{equation}
Here $h_{ij}$ is the 5d induced metric on a constant $t,r$ slice, namely
\begin{equation}
	h_{ij}=g_{\mu\nu}\fft{\partial x^\mu}{\partial x^i}\fft{dx^\nu}{\partial x^j}\bigg|_{t,r=\text{constant}}\,,
\end{equation}
and its determinant is given by
%
\begin{align}
	h&=\fft{y^2z^2(y^2-z^2)^2}{\Xi_1^2\Xi_2^2\Xi_3^2(a_1^2-a_2^2)^2(a_2^2-a_3^2)^2(a_3^2-a_1^2)^2}\notag\\
	&\quad\times\Bigg[\fft{\Xi_1\Xi_2\Xi_3r^2((r^2+y^2)(r^2+z^2)-r^2(r^2+g^2y^2z^2))}{(1-g^2y^2)(1-g^2z^2)}U(r)+\fft{\mS(r)}{r^2}\\
	&\quad-\Big((a_1^2+a_2^2+a_3^2+a_1^2a_2^2a_3^2g^4)r^4+(1-g^2r^2)((a_1^2a_2^2+a_2^2a_3^2+a_3^2a_1^2)r^2+a_1^2a_2^2a_3^2)\Big)U(r)\Bigg]\,.\notag
\end{align}
%
The functions $\mathcal{S}$ is defined as
\begin{align}
	\mS(r)&\equiv \prod_{I=1}^2\Big((r^2+a_1^2)(r^2+a_2^2)(r^2+a_3^2)+2ms_I^2(r^2-a_1a_2a_3g)\Big)\\
	&\quad+2mg(c_1-c_2)^2(r^2+a_1^2)(r^2+a_2^2)(r^2+a_3^2)(a_1+a_2a_3g)(a_2+a_3a_1g)(a_3+a_1a_2g)\,.\notag
\end{align}
The three angular momenta of the solution can be computed by employing the standard Komar integrals over the spatial $S^5$ at asymptotic infinity. For $J_1$ we find
\begin{equation}
\begin{split}
	J_1&=-\fft{1}{16\pi G_N}\int_{S^5}\star dK_1\\
	&=\fft{\pi^2m}{16G_N\Xi_1\prod_{j=1}^3\Xi_j}\Bigg[4a_1c_1c_2+4g(1-c_1c_2)(a_2+a_1a_3g)(a_3+a_1a_2g)\\
	&\quad+(c_1-c_2)^2\bigg(2a_2a_3g+a_1\Big(1+2\Xi_1-\sum_{j=1}^3\Xi_j\Big)\bigg)\bigg(1+2a_1a_2a_3g^3-\sum_{j=1}^3\Xi_j\bigg)\Bigg]\,,
\end{split}\label{eq:J}
\end{equation}
where $K_i$ is dual to a Killing vector $-\partial_{\phi_i}$, namely
\begin{equation}
	K_i^\mu\partial_\mu=-\partial_{\phi_i}\qquad\Leftrightarrow\qquad K_i=-g_{\mu\phi_i}dx^\mu\,.
\end{equation}
The other two angular momenta $J_{2,3}$ can be derived in a similar fashion and are given by an expression analogous to \eqref{eq:J} above obtained by an appropriate permutation of the parameters $a_i$.

The two electric charges  of the solution can also be computed using Komar integrals. For $Q_1$ we find 
\begin{equation}
\begin{split}
	Q_1&=-\fft{1}{16\pi G_N}\int_{S^5}(X_1^{-2}\star F^1_{(2)}-F^1_{(2)}\wedge A_{(3)})\\
	&=\fft{\pi^2 ms_1}{4G_N\Xi_1\Xi_2\Xi_3}\Big[2c_1-(c_1-c_2)g^2(a_1^2+a_2^2+a_3^2+2a_1a_2a_3g)\Big]\,.
\end{split}\label{eq:Q}
\end{equation}
The other electric charge $Q_2$ can be obtained in a similar manner and takes the same form as the expression above with $\{c_1,s_1\}$ exchanged with $\{c_2,s_2\}$. Due to the presence of Chern-Simons terms in the supergravity Lagrangian the definition of conserved and quantized charges is in general subtle, see \cite{Marolf:2000cb}. We have included the contribution from the Chern-Simons terms in the integrand in \eqref{eq:Q} but for the solution of interest this contribution vanishes sufficiently fast at asymptotic infinity and does not contribute to the electric charge. We thus conclude that the potentially different notions of charge coincide for the solution of interest. 

To determine the temperature of the black hole we work with Euclidean signature via the Wick rotation
\begin{equation}
	t=-\ri\tau\,,\label{Wick}
\end{equation}
and study the regularity of the metric near the outer horizon. To this end we can expand the metric in the near horizon region by defining a new radial coordinate $\rho$ as
\begin{equation}
	r=r_+ + \rho^2\,,
\end{equation}
and study the metric for small $\rho$ ($\rho^2\ll r_+$). The resulting Euclidean near horizon geometry is given from Appendix~\ref{App:equivalent} as
\begin{equation}
\begin{split}
	ds^2\big|_{\rho^2\ll r_+}&=(H_1H_2)^{\fft15}\fft{(r_+^2+y^2)(y^2-z^2)y^2}{(1-g^2y^2)(a_1^2-y^2)(a_2^2-y^2)(a_3^2-y^2)}dy^2\\
	&\quad+(H_1H_2)^{\fft15}\fft{(r_+^2+z^2)(z^2-y^2)z^2}{(1-g^2z^2)(a_1^2-z^2)(a_2^2-z^2)(a_3^2-z^2)}dz^2\\
	&\quad+(H_1H_2)^{\fft15}(r_+^2+y^2)(r_+^2+z^2)\bigg[\fft{r_+^4}{\mS(r_+)}U'(r_+)\rho^2d\tau^2+\fft{4d\rho^2}{U'(r_+)}\bigg]\\
	&\quad+\sum_{i,j=1}^3g_{ij}(r_+,y,z)(d\phi_i+\ri\Omega_id\tau)(d\phi_j+\ri\Omega_jd\tau)\,,
\end{split}\label{metric:nH}
\end{equation}
under the Wick rotation \eqref{Wick}. The regularity of the near horizon geometry \eqref{metric:nH} at $\rho =0$ then determines the periodic identification of coordinates along the thermal cycle as
\begin{equation}\label{eq: bulk periodicity}
	(\tau,\phi_i)\sim(\tau+\beta,\phi_i-\ri\Omega_i\beta)\qquad\text{where}\qquad\beta=2\pi\fft{2\sqrt{\mS(r_+)}}{r_+^2U'(r_+)}\,.
\end{equation}
The Hawking temperature of the black hole is then identified with the inverse of $\beta$ and thus we find 
\begin{equation}
	T=\beta^{-1}=\fft{r_+^2U'(r_+)}{4\pi\sqrt{\mS(r_+)}}\,.\label{eq:T}
\end{equation}
The same result for the temperature can be obtained by computing the surface gravity at the outer horizon.

The angular velocities $\Omega_i$ of the black hole can be determined by finding the unique Killing vector
\begin{equation}
	\ell=\partial_t+\sum_{i=1}^3\Omega_i\partial_{\phi_i}\,,
\end{equation}
which is null at the outer horizon, i.e. it obeys $\ell^\mu\ell_\mu\big|_{r=r_+}=0$. One finds that $\Omega_1=w_1(r_+)$ where the function $w_1(r)$ is given by
\begin{align}
	w_1(r)&=\fft{1}{\mS(r)}\Bigg[\fft12\bigg(\prod_{i=1}^3(r^2+a_i^2)+2ms_1^2(r^2-a_1a_2a_3g)\bigg)\nn\\
	&~\times\bigg(a_1(1+g^2r^2)(r^2+a_2^2)(r^2+a_3^2)+2ms_2^2(a_1g^2r^2-a_2a_3g)\bigg) \notag\\
	& +\fft12\bigg(\prod_{i=1}^3(r^2+a_i^2)+2ms_2^2(r^2-a_1a_2a_3g)\bigg) \notag \\& \times  \bigg(a_1(1+g^2r^2)(r^2+a_2^2)(r^2+a_3^2)+2ms_1^2(a_1g^2r^2-a_2a_3g)\bigg) \notag \\
	&-m(c_1-c_2)^2(r^2+a_2^2)(r^2+a_3^2)\Big\{2g(gr^2-a_1)(a_1+a_2a_3g)(a_2+a_3a_1g)(a_3+a_1a_2g)\notag\\
	&+(1+a_1g)^2(a_1+a_2a_3g)(1-a_1g-a_2g-a_3g)(1-a_1g+a_2g+a_3g)r^2\Big\}\Bigg]\,.\label{eq:Omega}
\end{align}
The other two angular velocities $\Omega_{2,3}$ are determined by similar functions $w_{2,3}(r)$ evaluated at $r_{+}$ which can be obtained from the expression for $w_1(r)$ by employing the permutation symmetry. 


The electrostatic potentials of the solution can be computed by taking the difference
\begin{equation}
\Phi_I = \ell^\mu A^I_{(1)\mu}\Big|_{r=r_+} - \ell^\mu A^I_{(1)\mu}\Big|_{r \to \infty}\,. 
\end{equation}
The result for $\Phi_I$ does not depend on the pure gauge parameters $\alpha_I$ introduced in \eqref{eq:A1Idef} and we find that $\Phi_1$ is given by 
\begin{align}
	&\Phi_1=\fft{2mr_+^2s_1c_1\big(\prod_i(r_+^2+a_i^2)+2ms_2^2(r_+^2-a_1a_2a_3g)\big)}{\mS(r_+)}\notag\\
	&-\fft{mgr_+^2s_1(c_1-c_2)}{\mS(r_+)}\Big\{g(a_1^2+a_2^2+a_3^2+2a_1a_2a_3g)\big(\prod_i(r_+^2+a_i^2)+2ms_2^2(r_+^2-a_1a_2a_3g)\big)\notag\\
	&\kern10em+4ms_2^2(a_1+a_2a_3g)(a_2+a_3a_1g)(a_3+a_1a_2g)\Big\}\,.
\label{eq:Phi}
\end{align}
The other electrostatic potential $\Phi_2$ can be obtained from the expression above by using the permutation symmetry. Note however that we should also impose that the norm of the gauge fields $A^I_{(1)}$ is regular at the black hole horizon.\footnote{One might also consider an analogous requirement for the 3-form potential $A_{(3)}$, achieved by virtue of a pure gauge shift. We do not explore this requirement here.} It turns out that these norms diverge at the horizon unless we set the pure gauge parameters $\alpha_I$ to be equal to the electrostatic potentials, i.e. we find that regularity of the gauge fields at the horizon imposes
\begin{equation}\label{eq: alpha sol}
\alpha_I = -\Phi_I\,.
\end{equation}
To calculate the energy, or mass, of the black hole solution we use the ADM formalism adapted for asymptotically locally AdS backgrounds, see for instance \cite{Ashtekar:1999jx,Chen:2005zj,Chow:2011fh}. After defining a properly conformally rescaled metric $\bar g_{\mu\nu} = \widetilde{\Omega}^{2} g_{\mu\nu}$ the energy is given by
\begin{equation}\label{eq:EAMDdef}
		E=\frac{1}{8(D-3) \pi g^3} \int_{\mathcal{B}} d \mathcal{B}_a \widetilde{\Omega}^{-(D-3)} \bar{n}^c \bar{n}^d \bar{C}^a{}_{c b d} \xi^b\,,
\end{equation}
where $D$ is the number of space-time dimensions ($D=7$ in our setup), $d\mathcal{B}_{a}$ is the area element of the $S^{D-2}$ section of the conformal boundary, $\widetilde{\Omega}$ is the conformal factor, $\xi$ is the time-like Killing vector, $\bar n_{c} = \partial_{c}\widetilde{\Omega}$ and $C^{\mu}{}_{\nu\rho\sigma}$ is the Weyl tensor. Note that the Weyl tensor for $g_{\mu\nu}$ with the first index raised is equal to that for $\bar g_{\mu\nu}$, i.e., it is invariant under conformal rescaling of the metric.
	
For our solution, we can use the conformal factor
\begin{align}
	\widetilde{\Omega} = \frac{1}{g r}\,.
\end{align}
The only non-zero component of the time-like Killing vector is $\xi^{t} = 1$ and we have
\begin{align}
		g_{rr} &= \frac{1}{g^2 r^2}\,, \qquad g^{rr} = g^2 r^2\,,	\qquad \bar{C}^a{}_{c b d} = \mathcal{O}(r^{-8})\,,	 \notag	\\
		\bar n^{r} &= \bar{g}^{ra}\bar{n}_{a} = \bar g^{rr} \left(-\frac{1}{gr^2}\right) = \widetilde{\Omega}^{-2} g^{rr}\left(-\frac{1}{gr^2}\right) = -\widetilde{\Omega}^{-2} g\,.			
	\end{align}
The 6-volume element at the AdS boundary is given by
\begin{align}
	\begin{split}
	\operatorname{vol}_6 =
	\frac{y z \left|y^2-z^2\right|  dt \wedge dy\wedge dz\wedge d\phi_1 \wedge d\phi_2 \wedge d\phi_3}{g^5 \left(a_{1}^2-a_{2}^2\right) \left(a_{3}^2-a_{1}^2\right)\left(a_{2}^2-a_{3}^2\right) \Xi_1 \Xi_2 \Xi_3 } + \mathcal{O}(r^{-1})\,.
	\end{split}
\end{align}
Following \cite{Chen:2005zj}, we define $d\mathcal{B}_{\mu} \equiv \langle \partial_{\mu}, \operatorname{vol} \rangle$ where the angle brackets denote an inner product between a vector and a form defined by $\langle \partial_{\mu}, dx^{\nu} \rangle = \delta^{\mu}_{\nu}$. Applying this to the time-like Killing vector and volume 6-form above we find
\begin{align}
	d\mathcal{B}_{t} &= \frac{y z \left|y^2-z^2\right| dy\wedge dz\wedge d\phi_1 \wedge d\phi_2 \wedge d\phi_3}{g^5 \left(a_{1}^2-a_{2}^2\right) \left(a_{3}^2-a_{1}^2\right)\left(a_{2}^2-a_{3}^2\right) \Xi_1 \Xi_2 \Xi_3 }\,.
\end{align}
Using these results in the general expression for the ADM mass in \eqref{eq:EAMDdef} we find that the energy of our black hole solution is given by the integral
\begin{align}
	E &= \frac{1}{32\pi g}\int_{\mathcal{B}} d\mathcal{B}_{t} \widetilde{\Omega}^{-8}\bar{C}^{t}{}_{rtr}\,.
\end{align}
We refrain from presenting the explicit expression for $\bar{C}^{t}{}_{rtr}$ because it is not very illuminating. The integral is taken over the coordinates $\{y,z,\phi_1,\phi_2,\phi_3\}$ whose ranges are discussed around \eqref{eq:yandzrange}. Performing this integral we find the following explicit expression for the energy of the black hole solution
\begin{align}
	E&=\fft{m\pi^2}{8G_N\Xi_1\Xi_2\Xi_3}\Bigg[\sum_{i=1}^3\fft{2}{\Xi_i}-1+\fft{5(s_1^2+s_2^2)}{2}+\fft{s_1^2+s_2^2}{2}\sum_{i=1}^3\bigg(\fft{2(1+a_i^2g^2-\Sigma_2-2\Pi_1)}{\Xi_i}-\Xi_i\bigg)\Bigg] \notag\\
	&+\fft{m\pi^2(c_1-c_2)^2}{32G_N\Xi_1^2\Xi_2^2\Xi_3^2}\Bigg[-10\Sigma_2-16\Pi_1+11\Sigma_4+13\Pi_{22}+32\Pi_1\Sigma_2-3(\Sigma_6+5\Pi_{42}+4\Pi_1^2)\notag\\
	&\hspace{13mm}-16\Pi_1\Sigma_2^2+\Pi_{62}+3\Pi_{44}-5\Pi_1^2\Sigma_2+8\Pi_1(2\Pi_2+\Pi_{42})+\Pi_1^2(\Sigma_4+3\Pi_{22})\Bigg]\,,
\label{eq:E}
\end{align}
where we have defined the parameters 
\begin{align}
	\Sigma_n&\equiv(a_1g)^n+(a_2g)^n+(a_3g)^n\,,\notag\\
	\Pi_n&\equiv(a_1g)^n(a_2g)^n(a_3g)^n\,,\label{SigmaPi}\\
	\Pi_{nm}&\equiv(a_1g)^n((a_2g)^m+(a_3g)^m)+(a_2g)^n((a_3g)^m+(a_1g)^m)+(a_3g)^n((a_1g)^m+(a_2g)^m)\,.\notag
\end{align}
The expressions we derived for the various thermodynamic quantities of the black hole are unwieldy and it is therefore important to have a non-trivial consistency check of their validity. The first law of black hole thermodynamics offers precisely such a consistency check. To implement the first law we take the expressions for $E$ \eqref{eq:E}, $T$ \eqref{eq:T}, $S$ \eqref{eq:S}, $\Omega_{i}$ \eqref{eq:Omega}, $J_{i}$ \eqref{eq:J}, $\Phi_{I}$ \eqref{eq:Phi} and $Q_{I}$ \eqref{eq:Q} and vary them with respect to the parameters specifying the black hole solution $(m,a_i,\delta_I)$. After performing this arduous calculation we find that indeed the first law of black hole thermodynamics is obeyed, i.e. 
\begin{equation}
	dE=TdS+\sum_{i=1}^3\Omega_i dJ_i+\sum_{I=1}^2\Phi_I dQ_I\,.
\end{equation}
An important observable in Euclidean quantum gravity with great significance in holography is the regularized on-shell action of any smooth asymptotically AdS classical solution. The proper calculation of the on-shell action requires a careful implementation of the holographic renormalization procedure. In the context of stationary AdS black hole solutions these types of calculations are notoriously subtle especially if one is interested in taking supersymmetric or extremal limits of these solutions, see for instance \cite{Cassani:2019mms} for a recent account. In lieu of a proper on-shell action calculation we proceed with evaluating the regularized on-shell action of the black hole solution of interest by employing the so called quantum statistical relation \cite{Gibbons:1976ue} which can be viewed as an integrated form of the first law of black hole thermodynamics. For the solutions of interest we therefore find the following long expression for the on-shell action
\begin{align}
	I&=-S+\beta\Bigg[E-\sum_{i=1}^3\Omega_iJ_i-\sum_{I=1}^2\Phi_IQ_I\Bigg] \notag\\
	&=\fft{\pi^2\beta}{16G_N\Xi_1\Xi_2\Xi_3r_+^2}\Bigg[(1-g^2r_+^2){\textstyle\prod_{i=1}^{3}}(r_+^2+a_i^2)-2m(s_1^2+s_2^2)(g^2r_+^4+a_1a_2a_3g)\notag\\
	&\quad-\fft{2m^2s_1^2s_2^2}{g^4\mS(r_+)}\Big({\textstyle\prod_{i=1}^{3}}(r_+^2+a_i^2)+2ms_1^2(r_+^2-a_1a_2a_3g)\Big)\notag\\
	&\qquad\times\Big(g^6r_+^6+(\Sigma_2+2\Pi_1)g^4r_+^4-(2\Pi_1+\fft12\Pi_{22}-2g^4ms_2^2)g^2r_+^2+\Pi_1(2g^4ms_2^2-\Pi_1)\Big)\notag\\
	&\quad-\fft{2m^2s_1^2s_2^2}{g^4\mS(r_+)}\Big({\textstyle\prod_{i=1}^{3}}(r_+^2+a_i^2)+2ms_2^2(r_+^2-a_1a_2a_3g)\Big)\notag\\
	&\qquad\times\Big(g^6r_+^6+(\Sigma_2+2\Pi_1)g^4r_+^4-(2\Pi_1+\fft12\Pi_{22}-2g^4ms_1^2)g^2r_+^2+\Pi_1(2g^4ms_1^2-\Pi_1)\Big)\notag\\
	&\quad+\fft{m(c_1-c_2)^2(a_1+a_2a_3g)(a_2+a_3a_1g)(a_3+a_1a_2g)}{g^{11}\mS(r_+)}\notag\\
	&\quad\times\bigg(-2\Xi_1\Xi_2\Xi_3\Big(g^6r_+^6+\Sigma_2g^4r_+^4+(\tfrac{1}{2}\Pi_{22}+2m g^4(s_1^2+s_2^2))g^2r_+^2+\Pi_2\notag\\
	&\qquad-2g^4m(1+s_1^2+s_2^2)-\fft12(c_1-c_2)^2g^4m(-2\Sigma_2-8\Pi_1+\Sigma_4-\Pi_{22})+(s_1^2+s_2^2)g^4m\Sigma_2\Big)\notag\\
	&\qquad+g^4m\Big(4+8g^4ms_1^2s_2^2+(2-\Sigma_2-2\Pi_1)(2(s_1^2+s_2^2)-(c_1-c_2)^2(\Sigma_2+2\Pi_1))\Big)\notag\\
	&\qquad\times\fft{g^8r_+^2{\textstyle\prod_{i=1}^{3}}(r_+^2+a_i^2)-\Xi_1\Xi_2\Xi_3}{1+g^2r_+^2}-8g^8m^2s_1^2s_2^2(2g^6r_+^6+\Sigma_2g^4r_+^4-\Pi_2)\bigg)\Bigg]\,.
\label{eq:nonBPSI}
\end{align}
We leave the derivation of this result using the tools of holographic renormalization for future work.

\subsection{Supersymmetric limits}\label{sec:analysis:susy}

It is natural to investigate whether there are limits of the supergravity solution presented that are invariant under part of the supersymmetry transformations of the 7d maximal gauged supergravity theory. To study this question properly requires a careful analysis of the supersymmetry variations of the supergravity theory and the construction of the corresponding Killing spinors. Unfortunately this is a technically difficult task. We can circumvent this technical obstacle by employing a trick often used when analyzing the supersymmetry properties of black hole solutions in supergravity, see for instance \cite{Cvetic:2005zi,Chow:2007ts} for a more detailed discussion. The idea is to exploit the fact that a supersymmetric solution results in a linear relation between energy, angular momenta and electric charges dictated by the supersymmetry algebra and to find a relation between the parameters that determine the supergravity solution that ensures this linear relation. Importantly, this supersymmetric limit can be applied to the Euclidean supergravity solution at hand and is logically distinct from the extremal limit, also called BPS limit, of vanishing temperature which is appropriate for the Lorentzian black hole.

Guided by the results in \cite{Cvetic:2005zi,Chow:2007ts} we find that the Euclidean supersymmetric limit is achieved by imposing the following relation between the parameters specifying the supergravity solution
\begin{equation}
	(a_1+a_2+a_3)g=\fft{2}{1-e^{\delta_1+\delta_2}}\,.\label{susy}
\end{equation}
Note that the parameter $m$ is not constrained and we thus have a 5-parameter family of supersymmetric regular Euclidean solutions. For general non-vanishing values of the charges and parameters that specify the Euclidean supergravity solution we expect that it preserves two real supercharges.\footnote{It is possible that there is supersymmetry enhancement for specific values of the parameters $\{m,a_i,\delta_I\}$ but we do not study this further here.}

Indeed, imposing \eqref{susy} we find that the angular momenta (\ref{eq:J}), the electric charges (\ref{eq:Q}), and the energy (\ref{eq:E}) satisfy the relation
\begin{equation}\label{eq:susylinearcharges}
	E+g\sum_{i=1}^3J_i-\sum_{I=1}^2Q_I=0\,.
\end{equation}
The supersymmetric limit in \eqref{susy} also leads to a linear relation between the angular velocities and electrostatic  potentials of the solution that reads
\begin{equation}\label{eq:OmegaPhi2ndsheet}
	\beta\Bigg[g-\sum_{i=1}^3\Omega_i-2g\sum_{I=1}^2\Phi_I\Bigg]=\pm2\pi\ri\,.
\end{equation}
This relation deserves a few comments. First, notice that since the right hand side is imaginary this necessarily means that some of the quantities on the left hand side are generically complex.\footnote{The choice of sign on the right hand side is not important for the discussion below and arises from the choice of root of the algebraic equation that determines the location of the outer horizon $U(r_{+})=0$. We solve this equation for the parameter $m$ which leads to two complex conjugated roots that correspond to the two signs in \eqref{eq:OmegaPhi2ndsheet}. } This seems problematic for Lorentzian gravitational solutions but is allowed for Euclidean ones. In fact, relations analogous to \eqref{eq:OmegaPhi2ndsheet} arise in the supersymmetric limits of Euclidean asymptotically AdS black holes in four and five dimensions as recently discussed in \cite{Cabo-Bizet:2018ehj,Cassani:2019mms,Bobev:2019zmz}. The linear relation in \eqref{eq:OmegaPhi2ndsheet} reflects a similar identity between the fugacities in the dual SCFT as we discuss further in the next section. 

The utility of the supersymmetric limit of the Euclidean solution presented above is that the regularized on-shell action in \eqref{eq:nonBPSI} remains finite and well-behaved. After imposing the relation in \eqref{susy} on the black hole parameters we find that the unwieldy expression in \eqref{eq:nonBPSI} can be written compactly as
\begin{equation}\label{eq:Ionshellsusy}
	I=-S+\beta\Bigg[E-\sum_{i=1}^3\Omega_iJ_i-\sum_{I=1}^2\Phi_IQ_I\Bigg]=\fft{\pi^2}{8G_Ng}\fft{\varphi_1^2\varphi_2^2}{\omega_1\omega_2\omega_3}\,,
\end{equation}
where we have defined
\begin{equation}\label{eq:varphiomega}
	\varphi_I\equiv\beta(\Phi_I-1)\,, \qquad\qquad \omega_i\equiv\beta(\Omega_i+g)\,.
\end{equation}
For general values of the six black hole parameters subject to the supersymmetric constraint in \eqref{susy} the Euclidean supergravity solution does not admit a ``good'' analytic continuation to Lorentzian signature. By this we mean that the analytic continuation introduces some kind of a pathology in the solution, manifested in terms of complex background fields or some form of causal pathology like closed time-like curves. It is however possible to impose an additional constraint on the black hole parameters that leads to a family of proper Lorentzian supersymmetric backgrounds. This additional constraint is equivalent to imposing an extremal limit to the non-supersymmetric Lorentzian black hole, i.e. to setting 
\begin{equation}
	T=0\quad\Leftrightarrow\quad U'(r_+)=0\,.\label{ext}
\end{equation}
We refer to the limit in which we impose both supersymmetry, \eqref{susy} and extremality \eqref{ext} as the BPS limit. In the BPS limit there are only four free parameters that specify the solution. They can be specified by taking the five parameters $\{a_i,\delta_I\}$ to obey \eqref{susy} and additionally impose that $m$ is given by
\begin{align}
	m^\star&=\fft{M_1}{g^4M_2}\,,\nn\\
	M_1&=2(2\Sigma_1-2\Sigma_2-\Pi_{11}+2\Pi_{21}+2\Pi_1)(2\Pi_1+\Pi_{21})\prod_{i=1}^3(1-a_ig)\,,\label{eq:BPSmconstr}\nn\\
	M_2&=2(1-\Sigma_1)^3(2-(\Sigma_2+2\Pi_{11})(s_1^2+s_2^2))-8(s_1^2+s_2^2)(1-\Sigma_1)^2(2\Pi_1+\Pi_{21})\nn\\
	&\quad +(c_1-c_2)^2(1-\Sigma_1)^2(-2\Sigma_2+2\Sigma_1\Sigma_2-8\Pi_1+\Sigma_4-\Pi_{22}+8\Sigma_1\Pi_1 \nn\\&\quad-\Sigma_1^2(\Sigma_3-\Pi_{21}+2\Pi_1))\,,
\end{align}
where we use the parameters defined in \eqref{SigmaPi} and denote the BPS value of a parameter with a ``$\star$'' superscript. For instance for the BPS value of $r_{+}$ we find
\begin{equation}
r_+^\star=\sqrt{\fft{a_1a_2+a_2a_3+a_3a_1-a_1a_2a_3g}{1-(a_1+a_2+a_3)g}}\,,
\end{equation}
while the angular velocities (\ref{eq:Omega}) and the electrostatic potentials (\ref{eq:Phi}) take the BPS values 
\begin{equation}
	\Omega_i^\star=-g\,,\qquad\Phi_I^\star=1\,.
\end{equation}
Importantly, the quantities defined in \eqref{eq:varphiomega} remain finite in the BPS limit and the on-shell action takes the same functional form as in \eqref{eq:Ionshellsusy}.

The BPS constraint in \eqref{eq:BPSmconstr} can be rewritten in terms of the angular momenta and charges of the black hole solution and leads to the following non-linear constraint between them
\begin{equation}
\begin{split}
	&\fft{\fft{2}{g^3}Q_1^\star Q_2^\star(Q_1^\star+Q_2^\star)-\fft{\pi^2}{G_Ng^5}(J_1^\star J_2^\star+J_2^\star J_3^\star+J_3^\star J_1^\star)}{\fft{8}{g}(Q_1^\star+Q_2^\star)-\fft{\pi^2}{G_Ng^5}}\\
	&=\bigg(-\fft{\pi^2}{16G_Ng^5}(J_1^\star+J_2^\star+J_3^\star)+\fft{1}{8g^2}((Q_1^\star)^2+(Q_2^\star)^2)+\fft{1}{2g^2}Q_1^\star Q_2^\star\bigg)\\
	&\quad\times\Bigg(1-\sqrt{1-\fft{-\fft{\pi^2}{8G_Ng^5}J_1^\star J_2^\star J_3^\star+\fft{1}{16g^4}(Q_1^\star)^2(Q_2^\star)^2}{\big(-\fft{\pi^2}{16G_Ng^5}(J_1^\star+J_2^\star+J_3^\star)+\fft{1}{8g^2}((Q_1^\star)^2+(Q_2^\star)^2)+\fft{1}{2g^2}Q_1^\star Q_2^\star\big)^2}}\Bigg)\,.
\end{split}\label{eq:charge:BPS}
\end{equation}
Similar non-linear relations between the angular momenta and charges of BPS black hole solutions in AdS have been observed also in four- and five-dimensional gauged supergravity, see \cite{Cassani:2019mms} for a summary. One of the consequences of the BPS constraint in \eqref{eq:BPSmconstr}, or equivalently \eqref{eq:charge:BPS}, is that the entropy of the Lorentzian BPS black hole is real and takes the following compact form
\begin{equation}\label{eq:BPSentropy}
	S^\star=2\pi\sqrt{\fft{\fft{2}{g^3}Q_1^\star Q_2^\star(Q_1^\star+Q_2^\star)-\fft{\pi^2}{G_Ng^5}(J_1^\star J_2^\star+J_2^\star J_3^\star+J_3^\star J_1^\star)}{\fft{8}{g}(Q_1^\star+Q_2^\star)-\fft{\pi^2}{G_Ng^5}}}\,.
\end{equation}
In summary, we have shown that there is a five-parameter family of regular supersymmetric Euclidean solutions parametrized by $\{m,a_i,\delta_I\}$ subject to the relation in \eqref{susy}. These backgrounds are asymptotically locally AdS$_7$ with an $S^1\times S^5$ spatial boundary, for which the $S^1$ smoothly shrinks to zero size at $r=r_+$. The angular momenta and charges of these solutions obey the linear relation in \eqref{eq:susylinearcharges}, while the $S^1$ circumference, $\beta$, the angular velocities and the electrostatic potentials obey \eqref{eq:OmegaPhi2ndsheet}. The on-shell action of these supersymmetric solution is finite and is given by \eqref{eq:Ionshellsusy}. To obtain regular supersymmetric Lorentzian black holes without causal pathologies we have to impose a further constraint on the black hole parameters which can be written in a number of equivalent ways, for instance as in \eqref{eq:BPSmconstr} or in terms of a non-linear relation between the black hole charges and angular momenta \eqref{eq:charge:BPS}. Imposing this constraint leads to a four-parameter family of supersymmetric and extremal Lorentzian black hole solutions with real entropy and finite on-shell action. In the next section we proceed with a discussion on a holographic interpretation of this family of supersymmetric solutions.

\section{Holography}
\label{sec:holo}

The general non-extremal black hole solution presented in Section~\ref{sec:BHsol} can be uplifted to 11d supergravity and should be holographically dual to the large $N$ limit of the 6d $\mathcal{N}=(2,0)$ SCFT at finite temperature and finite rotational and electric chemical potentials.  It is hard to check the AdS/CFT correspondence quantitatively in this setting since we do not have field theoretic tools to calculate the thermal partition function and other thermodynamic observables of the 6d $\mathcal{N}=(2,0)$ SCFT. As often happens, supersymmetry offers better technical control. The five-parameter family of Euclidean supersymmetric solutions described in Section~\ref{sec:analysis:susy} should be described holographically by the 6d $\mathcal{N}=(2,0)$ SCFT on $S^1\times S^5$ with appropriate background fields and periodicity conditions for the fermions on the circle. This supersymmetric path integral is often referred to as the ``superconformal index'' and was defined in \cite{Bhattacharya:2008zy} for 6d SCFTs.

In the holographic context we are interested in the leading term in the large $N$ limit of the logarithm of the $Z_{S^1\times S^5}$ path integral which should be dual to the regularized on-shell action in \eqref{eq:Ionshellsusy}. To the best of our knowledge this path integral has not been computed very rigorously in the large $N$ limit. There are however explicit expressions in the literature that are either conjectured or partially derived in certain limits. As summarized in \cite{Hosseini:2018dob,Choi:2018hmj,Nahmgoong:2019hko,Cassani:2019mms,Ohmori:2020wpk} it is expected that the large $N$ limit of $\log Z_{S^1\times S^5}$ scales as $N^3$ and is given by
\begin{equation}\label{eq:logZS1S5}
	-\log Z_{S^1\times S^5} = \frac{N^3}{24} \frac{\Delta_1^2\Delta_2^2}{\hat{\omega}_1\hat{\omega}_2\hat{\omega}_3}\,,
\end{equation}
where we have followed the convention of \cite{Choi:2018hmj}\footnote{Note, the Cartan generators $\{\hat{E}, \hat{J}_i, \hat{Q}_I\}$ used in \cite{Choi:2018hmj} are related to our bulk quantities $\{E, J_i, Q_I\}$ by $E=g\hat{E},J_i = -\hat{J}_i, Q_I = 2g \hat{Q}_I$.}. This expression includes the contribution of the supersymmetric Casimir energy of the 6d $\mathcal{N}=(2,0)$ SCFT on $S^1\times S^5$ derived in \cite{Bobev:2015kza} and can be viewed as a 6d generalization of the ``index on the second sheet'' derived for 4d $\mathcal{N}=1$ SCFTs in \cite{GonzalezLezcano:2020yeb,Cassani:2021fyv,ArabiArdehali:2021nsx}. The parameters $\hat{\omega}_i$ specify the twisting of the Euclidean time circle along $S^5$, while the $\Delta_I$ fix the value of a background R-symmetry flat connection. They satisfy the linear relation\footnote{More generally, we can consider $2\pi \ri(1+2n)$ for any $n\in \mathbb{Z}$ on the right-hand-side, but this shift can always be absorbed into some $\hat{\omega}_i$, since the index is periodic under $\hat{\omega}_i\sim \hat{\omega}_i + 4\pi \ri$. }
\begin{equation}\label{eq:Deltavarphi}
	\Delta_1 + \Delta_2 - \hat{\omega}_1 - \hat{\omega}_2 - \hat{\omega}_3 = \pm 2\pi \ri \, .
\end{equation}
To compare the field theory expression in \eqref{eq:logZS1S5} with the gravitational on-shell action in \eqref{eq:Ionshellsusy}, first we need to employ the holographic dictionary. The relation between the 7d Newton constant $G_N$, the AdS$_7$ scale $L=1/g$, and the number $N$ of M5-branes can be worked out by uplifting the AdS$_7$ vacuum solution of the 7d gauged supergravity theory to 11d and employing the usual flux quantization of 11d 4-form flux on the internal $S^4$. This leads to the following relation
\begin{equation}\label{eq:Ntogmap}
	N^3=\fft{3\pi^2}{16G_Ng^5}\,.
\end{equation}
Given the SCFT background described above, we can then seek the bulk Euclidean solution that dominates the gravitational path integral formulated with these asymptotics. This is achieved by studying regularity of the metric and gauge fields near the horizon, which precisely determines the relationship between bulk and boundary chemical potentials. From (\ref{eq: bulk periodicity}) and using the coordinate transformation (\ref{eq: spheroidal coord change}) we identify
\begin{align}\label{eq: AM matching}
  \hat{\omega}_i = \beta(\Omega_i + g) = \omega_i\,,
\end{align}
where we first introduced $\omega_i$ in (\ref{eq:varphiomega}). The non-normalizable modes of the bulk gauge fields $A_I$ specify the vacuum expectation values of the SCFT background R-symmetry gauge fields $\hat{A}_I$, subject to the normalization match $2g A_I \leftrightarrow \hat{A}_I$. Thus, from (\ref{eq: alpha sol}) we identify
\begin{align}\label{eq: R-symm matching}
  \Delta_I = -2g \beta(\Phi_I - 1)  = -2g \varphi_I\,,
\end{align}
with $\varphi_I$ as introduced in (\ref{eq:varphiomega}). We then see that the constraint (\ref{eq:Deltavarphi}) is equivalent to the bulk constraint (\ref{eq:OmegaPhi2ndsheet}) specifying the supersymmetric solution. Happily, we find that the large $N$ SCFT result (\ref{eq:logZS1S5}) does indeed match the gravitational on-shell action of this supersymmetric solution (\ref{eq:Ionshellsusy}), as using (\ref{eq:Ntogmap}), (\ref{eq: AM matching}) and (\ref{eq: R-symm matching}) we find
\begin{equation}
	-\log Z_{S^1\times S^5} = \frac{N^3}{24} \frac{\Delta_1^2\Delta_2^2}{\hat{\omega}_1\hat{\omega}_2\hat{\omega}_3} = \fft{\pi^2}{8G_Ng}\fft{\varphi_1^2\varphi_2^2}{\omega_1\omega_2\omega_3}\,.
\end{equation}
We consider this to be a non-trivial precision test of holography. We hasten to add that while the supergravity result in \eqref{eq:Ionshellsusy} is derived rigorously and is valid for all values of the angular velocities and electrostatic potentials, the derivation of \eqref{eq:logZS1S5} is most solidly established in the so-called Cardy-like limit of small $\hat{\omega}_i$. Motivated by our supergravity calculation of the on-shell action it will be interesting to revisit the analysis in \cite{Hosseini:2018dob,Choi:2018hmj,Ohmori:2020wpk,Nahmgoong:2019hko} and show that the results in \eqref{eq:logZS1S5} is valid for all values of the fugacities.

An alternative point of view of the path integral of the 6d $\mathcal{N}=(2,0)$ SCFT on $S^1\times S^5$ is by formulating it as an index that counts supersymmetric states. In a holographic context this then implies that the large $N$ behavior of the index may account for the entropy of the supersymmetric Lorentzian black holes that we presented in Section~\ref{sec:analysis:susy}. Although the exact BPS degeneracies are not accessible from the index, one can nonetheless study its coefficients in a Laurent expansion, which contain information on certain alternating sums over degeneracies. Such coefficients can be extracted by contour integral, which by the saddle-point approximation at large $N$ reduces to a constrained Legendre transform \cite{Hosseini:2018dob,Choi:2018hmj,Nahmgoong:2019hko}. The result, derived in \cite{Nahmgoong:2019hko} (see also \cite{David:2020ems}) for general angular momenta and charges, precisely agrees with the entropy of the supersymmetric Lorentzian black hole that we derived in \eqref{eq:BPSentropy}, and thus supports the expectation that the index is dominated by the 4-parameter Lorentzian BPS solution presented here. This analysis is subtle, due to the difficulties in computing the large $N$ behavior of the superconformal index rigorously, as well as the mysterious QFT nature of the non-linear constraints between the charges in \eqref{eq:charge:BPS}. Nevertheless, we view this analysis as a remarkable agreement between the supergravity calculation of the black hole entropy and its field theory counterpart.

Finally we note that in our conventions and after using \eqref{eq:Ntogmap} the black hole charges and angular momenta have the following scaling with the number of M5-branes
\begin{equation}
J_{i} \sim N^3\,, \qquad\qquad \frac{Q_I}{g} \sim N^3\,.
\end{equation}
Using this in \eqref{eq:BPSentropy} we indeed find that the black hole entropy scales as $N^3$ which is the expected M5-brane scaling behavior.

%
\section{Discussion}
\label{sec:discussion}

In this work we presented a new black hole solution of 7d maximal $\SO(5)$ gauged supergravity with three independent angular momenta and two electric charges. The uplift of this solution to 11d supergravity describes the backreaction of a large number $N$ of coincident M5-branes. We delineated the thermal properties of this black hole solution and studied its supersymmetric and extremal limits. We obtained a 5-parameter family of supersymmetric Euclidean solutions that have a finite on-shell action which obeys the quantum statistical relation and is in perfect agreement with the QFT calculations of the $S^1\times S^5$ path integral of the 6d $\mathcal{N}=(2,0)$ $A_{N-1}$ SCFT. A Legendre transformation of this result to the microcanonical ensemble of fixed electric charges and angular momenta leads to a supersymmetric black hole entropy in perfect agreement with the field theory results. 

Our results lead to several natural directions for generalizations and further work which we briefly outline below.

\begin{itemize}

\item It is easy to generalize the results above to the $D_N$ series of $\mathcal{N}=(2,0)$ SCFTs in the large $N$ limit. The 11d black hole supergravity solution remains the same with the minor replacement of the internal $S^4$ with its smooth $\mathbb{Z}_2$ orbifold, i.e. $\mathbb{RP}^4$. This in turn changes the volume of the internal space and introduces factors of $2$ in the M5-brane charge quantization calculation. When the dust settles we are left with the following identification between the number of M5-branes and the 7d gravitational parameters
\begin{equation}
	N^3=\fft{3\pi^2}{64G_Ng^5}\,.
\end{equation}
This minor modification of \eqref{eq:Ntogmap}, together with the result for the on-shell action in \eqref{eq:Ionshellsusy} and the definitions in Section~\ref{sec:holo}, leads to the following holographic prediction for the leading term in the large $N$ expansion of the $S^1\times S^5$ path integral of the 6d $\mathcal{N}=(2,0)$ $D_{N}$ SCFT
\begin{equation}
-\log Z_{S^1\times S^5} = \frac{N^3}{6} \frac{\Delta_1^2\Delta_2^2}{\hat{\omega}_1\hat{\omega}_2\hat{\omega}_3}\,.
\end{equation}
This result is in perfect agreement with the field theory analysis in \cite{Nahmgoong:2019hko}.


\item It will be interesting to understand whether there are more general black hole solutions, or analogous Euclidean supergravity backgrounds, than the ones presented here. The field theory supersymmetric localization results in \cite{Qiu:2013pta} suggest that there should be supersymmetric gravitational solutions with $S^1\times$SE$_5$ boundary topology where SE$_5$ is a 5d Sasaki-Einstein manifold. It will be interesting to construct these solutions explicitly and understand their properties. It will also be interesting to understand whether the black holes with $S^5$ horizons we constructed here are the most general rotating electrically charged regular AdS$_7$ black holes in 11d supergravity or there are more general black hole solutions, perhaps with scalar hair, that can be found. We are not aware of any QFT or holographic arguments for the existence of such hairy black holes and it is therefore important to settle this uniqueness question rigorously.

\item There are two straightforward but tedious technical calculations that we have not performed in this work. The first one is the uplift of the 7d black hole solution we presented to 11d. This can be implemented by using the uplift formulae of \cite{Nastase:1999cb,Nastase:1999kf} but due to the complexity of the 7d black hole solution we expect that the 11d supergravity background will be very complicated and we do not see compelling reasons to work it out explicitly. The other technical analysis we did not perform is the proper 7d calculation of the regularized on-shell action of the Euclidean black hole solution. This calculation is both arduous and subtle due to the presence of finite counterterms that we expect will be needed in the holographic renormalization procedure. Indeed, as shown in \cite{Kallen:2012zn,Minahan:2013jwa}, the on-shell action of global AdS$_7$ computed with the standard holographic renormalization counterterms does not agree with the dual QFT path integral. This was remedied in \cite{Bobev:2018ugk,Bobev:2019bvq} by employing finite counterterms that break the bulk gauge invariance. We expect similar counterterms to play a role in the proper evaluation of the on-shell action of the AdS$_7$ black hole presented above and we hope to pursue this analysis in the near future.

\item It will be interesting to study higher-derivative corrections to the 7d gauged supergravity theory and how they affect the black hole solution above and its on-shell action. This question is motivated by the AdS/CFT correspondence since there are predictions for the subleading term of order $N$ in the $S^1\times S^5$ path integral of the 6d $\mathcal{N}=(2,0)$ $A_{N-1}$ SCFT \cite{Nahmgoong:2019hko,Ohmori:2020wpk}. Deriving this order $N$ term using a supergravity calculation will be a highly nontrivial test of holography. Similar calculations were successfully performed for asymptotically AdS black holes in 4d and 5d gauged supergravity in \cite{Bobev:2020egg,Bobev:2020zov,Bobev:2021oku,Bobev:2022bjm,Cassani:2022lrk,Cassani:2023vsa}.

\item It was pointed out in \cite{Larsen:1997ge,Castro:2012av} that for many stationary black hole solutions the product of the areas of all inner and outer horizons is independent of the ADM mass. We have checked explicitly that this property is also true for the AdS$_7$ black hole solution presented here. Given the complexity of the metric functions and the mass, angular momenta, and charges of the black hole we resorted to numerics to perform this check. While this amounts to non-trivial additional evidence that the observation of \cite{Larsen:1997ge,Castro:2012av} holds for general stationary black holes in various dimensions, the physical reason for its validity remains mysterious.

\item We suspect that there are generalizations of the black hole solution we presented above which can be found in 7d half-maximal gauged supergravity coupled to matter multiplets. We hope that it is possible to study the general matter coupled 7d half-maximal gauged supergravity carefully and using an Ansatz inspired by our results above construct the most general black hole solution in this theory. We suspect that this is a solvable problem since the structure of 7d gauged supergravity theories is quite constrained, see \cite{Louis:2015mka} for a review and a list of references. The additional requirement that the 7d theory admits a supersymmetric AdS$_7$ vacuum solution, as dictated by holography, significantly constrains the possible 7d supergravity actions. If such a general black hole solution can indeed be constructed it should be possible to compute its on-shell action. The QFT results in \cite{Nahmgoong:2019hko,Ohmori:2020wpk} strongly suggests that this on-shell action is fully controlled by the anomalies of the dual 6d $\mathcal{N}=(1,0)$ SCFT. It will be most interesting to study this further and we hope to do so in the near future.

\end{itemize}

\section*{Acknowledgments}

We are grateful to Pablo Cano, Davide Cassani, Alejandra Castro, Anthony Charles, Vasko Dimitrov, Robie Hennigar, Seyed Morteza Hosseini, Vincent Min, Valentin Reys, Paul Richmond, and Annelien Vekemans for useful discussions. NB is grateful to Mathew Bullimore and Hee-Cheol Kim for the enjoyable collaboration on~\cite{Bobev:2015kza} and numerous inspiring discussions about the superconformal index. This research is supported by the FWO projects G003523N and G094523N. NB, MD and JH are also supported in part by the KU Leuven C1 grant ZKD1118 C16/16/005 and by Odysseus grant G0F9516N from the FWO. MD is thankful for the hospitality of Benasque where the final stages of writing were completed. RM is supported by David Tong's Simons Investigator Award and is grateful to KU Leuven for warm hospitality in the early stages of this project.


\appendix

\section{Conventions}\label{App:convention}

In this section we summarize our conventions for differential forms and the Hodge star operation. 

Consider a differential form living in a $d$-dimensional manifold written in terms of tensor components as
\begin{equation}
	X_{(p)}=\fft{1}{p!}X_{\mu_1\cdots\mu_p}dx^{\mu_1}\wedge\cdots dx^{\mu_p}\qquad(0\leq p\leq d)\,,
\end{equation}
where the subscript $(p)$ denotes the rank of the form. The Hodge star operation is then defined as (we follow the convention given in section 7.9.2 of \cite{Nakahara:2003nw})
\begin{equation}
	\star(dx^{\mu_1}\wedge\cdots\wedge dx^{\mu_p})=\fft{\sqrt{|g|}}{(d-p)!}\varepsilon^{\mu_1\cdots\mu_p}{}_{\nu_1\cdots\nu_{d-p}}dx^{\nu_1}\wedge\cdots\wedge dx^{\nu_{d-p}}\,,\label{def:hodge}
\end{equation}
where the anti-symmetric symbol is given by
\begin{equation}
	\varepsilon_{\mu_1\cdots\mu_d}=\begin{cases}
		+1 & \text{if }\mu_1\cdots\mu_d\text{ is an even permutation of coordinate indices} \\
		-1 & \text{if }\mu_1\cdots\mu_d\text{ is an odd permutation of coordinate indices} \\
		0 & \text{otherwise}
	\end{cases}\,.
\end{equation}
Based on the definition (\ref{def:hodge}), it is easy to derive
\begin{equation}
\begin{split}
	\star\left(\star X_{(p)}\right)&=(-1)^{s+p(d-p)}X_{(p)}\,,\\
	X_{(p)}\wedge\star Y_{(p)}&=Y_{(p)}\wedge\star X_{(p)}\\
	&=\fft{1}{p!}X_{\mu_1\cdots\mu_p}Y^{\mu_1\cdots\mu_p}\sqrt{|g|}\,dx^1\wedge\cdots\wedge dx^d\,,
\end{split}
\end{equation}
where $s=1\,[0]$ for Lorentzian\,[Riemannian] manifolds respectively.

\section{Limits and comparison to the literature}
\label{App:limits}
The most general black hole solution we found in Section~\ref{sec:BHsol} reduces to three known solutions that have special values of the angular momentum and electric charge parameters, namely
\begin{equation}
	\begin{alignedat}{2}
		\text{i)}&~\text{3-rotations \& 2-equal-charges}&~:\quad &(a_1,a_2,a_3,\delta_1,\delta_2)=(a_1,a_2,a_3,\delta,\delta)\,,\\
		\text{ii)}&~\text{1-rotation \& 2-charges}&~:\quad &(a_1,a_2,a_3,\delta_1,\delta_2)=(0,0,a,\delta_1,\delta_2)\,,\\
		\text{iii)}&~\text{3-equal-rotations \& 2-charges}&~:\quad &(a_1,a_2,a_3,\delta_1,\delta_2)=(-a,-a,-a,\delta_1,\delta_2)\,,
	\end{alignedat}
\end{equation}
under the coordinate transformation \eqref{eq:yandzrange} introduced for cases ii) and iii) where some of the rotation parameters either vanish or are equal. In the following Subsections we provide more details on how to obtain these special limiting solutions. \\

One can also check that our black hole solution in the gauged supergravity reduces to a Kerr-Newman black hole solution in the ungauged supergravity under the limit $g\to0$, where the latter was obtained in \cite{Cvetic:1996dt} for general cases and then specialized for the case of our interest in Appendix A of \cite{Wu:2011gp}. Compared to reproducing the aforementioned three special gauged supergravity black hole solutions, it is more straightforward to reproduce the ungauged supergravity black hole solution under $g\to0$ so we do not provide further details below.\footnote{For comparison we took $\alpha_I=0$ in (\ref{eq:A1Idef}) and fixed the overall sign typo in the 2-form potential in Appendix A of \cite{Wu:2011gp}.}

\subsection{3-rotations \& 2-equal-charges}
We first summarize the 3-rotations \& 2-equal-charges black hole solution given in \cite{Chow:2007ts}. 

The metric reads
\begin{align}
		ds^2&=H^{\fft25}\left(-\fft{(1+g^2r^2)(1-g^2y^2)(1-g^2z^2)}{(1-a_1^2g^2)(1-a_2^2g^2)(1-a_3^2g^2)}dt^2
		+\fft{(r^2+y^2)(r^2+z^2)}{R}dr^2
		\right.\nn\\&\left.\quad+\fft{(r^2+y^2)(y^2-z^2)y^2}{(1-g^2y^2)(a_1^2-y^2)(a_2^2-y^2)(a_3^2-y^2)}dy^2
		\right.\nn\\&\left.\quad+\fft{(r^2+z^2)(z^2-y^2)z^2}{(1-g^2z^2)(a_1^2-z^2)(a_2^2-z^2)(a_3^2-z^2)}dz^2
		+\fft{(a_1^2+r^2)(a_1^2-y^2)(a_1^2-z^2)}{(a_1^2-a_2^2)(a_1^2-a_3^2)(1-a_1^2g^2)}d\phi_1^2
		\right.\nn\\&\left.\quad+\fft{(a_2^2+r^2)(a_2^2-y^2)(a_2^2-z^2)}{(a_2^2-a_1^2)(a_2^2-a_3^2)(1-a_2^2g^2)}d\phi_2^2
		+\fft{(a_3^2+r^2)(a_3^2-y^2)(a_3^2-z^2)}{(a_3^2-a_1^2)(a_3^2-a_2^2)(1-a_3^2g^2)}d\phi_3^2
		\right.\nn\\&\left.\quad
		+\fft{1-\fft{1}{H}}{H}\mA[y^2,z^2,0]\Big(C_0dt+C_1d\phi_1+C_2d\phi_2+C_3d\phi_3\Big)\,\right).\label{sol1:metric}
\end{align}
The 1-forms are given by
\begin{equation}
	\begin{split}
		A^I_{(1)}&=\fft{2msc}{H(r^2+y^2)(r^2+z^2)}\mA[y^2,z^2,0]=\bigg(1-\fft{1}{H}\bigg)\fft{c}{s}\mA[y^2,z^2,0]\,,\label{sol1:1form}
	\end{split}
\end{equation}
and the 3-form potential and 2-form potential are
\begin{align}
	\begin{split}
		A_{(3)}&=2ms^2g^4a_1a_2a_3\bigg[\mA[y^2,z^2,0]-\mA[y^2,z^2,g^{-2}]\bigg]\\
		&\qquad\wedge\bigg[\fft{dz\wedge\left(\mA[y^2,0,0]-\mA[y^2,0,g^{-2}]\right)}{(r^2+y^2)z}+\fft{dy\wedge\left(\mA[z^2,0,0]-\mA[z^2,0,g^{-2}]\right)}{(r^2+z^2)y}\bigg]\\
		&\quad+2ms^2g^3\mA[y^2,z^2,0]\\
		&\qquad\wedge\left[\fft{z\, dz\wedge\left(\mA[y^2,0,0]-\mA[y^2,0,g^{-2}]\right)}{(r^2+y^2)}+\fft{y\, dy\wedge\left(\mA[z^2,0,0]-\mA[z^2,0,g^{-2}]\right)}{(r^2+z^2)}\right]\,,\label{sol1:3form}
	\end{split}
	\\
	\begin{split}
		A_{(2)}&=\fft{2ms^2(a_1+a_2a_3g)}{H(r^2+y^2)(r^2+z^2)}\Bigg(\fft{(1-g^2y^2)(1-g^2z^2)\mu_1^2}{(1-a_1^2g^2)(1-a_2^2g^2)(1-a_3^2g^2)}dt\wedge d\phi_1\\
		&\kern12em+\fft{g(a_3^2-a_2^2)\mu_2^2\mu_3^2}{(1-a_2^2g^2)(1-a_3^2g^2)}d\phi_2\wedge d\phi_3\Bigg)\\
		&\quad+\text{(cyclic-permutations)}\,.\label{sol1:2form}
	\end{split}
\end{align}
In the above solution we have used the expressions \eqref{ansatz:functions}, \eqref{def:mu} and also introduced 
\begin{equation}
	\begin{split}
		s&=\sinh\delta\,,\quad c=\cosh\delta\,,\\
		R(r)&=\fft{1+g^2r^2}{r^2}(r^2+a_1^2)(r^2+a_2^2)(r^2+a_3^2)-2m+g^2r^2\left(r^2+\fft{2ms^2}{r^2}\right)^2-g^2r^6\\
		&\quad+2ms^2g^2(a_1^2+a_2^2+a_3^2)-\fft{4ms^2ga_1a_2a_3}{r^2}\,,\\
		C_0&=\fft{(1-g^2y^2)(1-g^2z^2)\big(c^2/s^2+(1-(a_1^2+a_2^2+a_3^2)g^2-2a_1a_2a_3g^3)H\big)}{(1-a_1^2g^2)(1-a_2^2g^2)(1-a_3^2g^2)}\,,\\
		C_i&=-\fft{a_i(a_i^2-y^2)(a_i^2-z^2)\big(c^2/s^2-(1-(a_i^2-a_j^2-a_k^2)g^2+\fft{2a_ja_k}{a_i}g)H\big)}{(a_i^2-a_j^2)(a_i^2-a_k^2)(1-a_i^2g^2)}\,,
	\end{split}
\end{equation}
where $i,j,k\in\{1,2,3\}$ are all different. In particular $C_0$ and $C_i$ are introduced to rewrite the metric in \cite{Chow:2007ts} for an easier comparison with the most general one \eqref{ansatz:metric}.

\medskip
\noindent\textbf{Comparison}
\medskip

\noindent 
Now it is straightforward to check that the most general black hole solution in Section~\ref{sec:BHsol} reduces to the 3-rotations \& 2-equal-charges solution described above under the identification
\begin{equation}
	\delta=\delta_1=\delta_2\,.\label{sol1:reduction}
\end{equation}
The thermodynamic quantities of the most general black hole solution evaluated in Section~\ref{sec:analysis:THD} reduce to the ones of the 3-rotations \& 2-equal-charges solution given in \cite{Chow:2007ts} as
\begin{equation}
	\begin{alignedat}{3}
		T&=T^\text{\cite{Chow:2007ts}}\,,&\qquad\Omega_i&=\Omega_i^\text{\cite{Chow:2007ts}}\,,&\qquad \Phi_I&=\Phi^\text{\cite{Chow:2007ts}}\,,\\
		S&=\fft{1}{G_N}S^\text{\cite{Chow:2007ts}}\,,&\qquad J_i&=\fft{1}{G_N}J_i^\text{\cite{Chow:2007ts}}\,,&\qquad Q_I&=\fft{1}{2G_N}Q^\text{\cite{Chow:2007ts}}\,,\\
		E&=\fft{1}{G_N}E^\text{\cite{Chow:2007ts}}\,,
	\end{alignedat}\label{general:to:case1}
\end{equation}
under the identification (\ref{sol1:reduction}) where we have restored the Newton constant explicitly.

\subsection{1-rotation \& 2-charges}
Next we consider the 1-rotation \& 2-charges black hole solution given in \cite{Chow:2011fh,Wu:2011gp}. Note that we follow the supergravity conventions of \cite{Chow:2007ts} summarized in Section~\ref{sec:7dsugra}, which differ from those of \cite{Chow:2011fh,Wu:2011gp} as follows
\begin{equation}
	\text{\cite{Chow:2007ts}}=\text{\cite{Wu:2011gp}}\big|_{g\to-g}=\text{\cite{Chow:2011fh}}\big|_{(g,A_{(3)})\to(-g,-A_{(3)})}\,.\label{case2:convention}
\end{equation}
Below we present the 1-rotation \& 2-charges solution consistent with the supergravity conventions in Section~\ref{sec:7dsugra} by flipping the signs of the gauge coupling or the 3-form potential in \cite{Chow:2011fh,Wu:2011gp} accordingly. For an easier comparison with the solution in Section~\ref{sec:BHsol}, we also rename coordinates as
\begin{equation}
	\begin{split}
		(t,r,y=a\sin\theta_1,\theta_2,\phi_1,\phi_2,\phi_3)&=(t,r,y=a\cos\theta,\psi,\xi,\zeta,\phi)^\text{\cite{Wu:2011gp}}\\
		&=(t,r,y,\vartheta,\phi_2,\phi_3,\phi)^\text{\cite{Chow:2011fh}}\,.
	\end{split}\label{sol2:coords}
\end{equation}
The metric is given by
\begin{equation}
	\begin{split}
		ds^2
		&=(H_1H_2)^{\fft15}\left(\fft{r^2+y^2}{\Delta_r}dr^2+\fft{r^2+y^2}{\Delta_y}dy^2+\fft{r^2y^2}{a^2}(d\theta_2^2+\sin^2\theta_2d\phi_1^2+\cos^2\theta_2d\phi_2^2) \right. \\
		&\left. \quad+\fft{1}{H_1H_2(1-a^2g^2)^2(r^2+y^2)}\left[-(V_y^2\Delta_r-\mV_1\mV_2\Delta_y)dt^2+(\mW_1\mW_2\Delta_y-\tV_y^2\Delta_r)a^2d\phi_3^2 \right. \right. \\
		& \left. \left. \kern14em~-\fft{4m'c'_1c'_2\tc'_1\tc'_2\Delta_y}{ar^2}dtd\phi_3\right]\right)\\
		&=(H_1H_2)^{\fft15}\left(-\fft{(1+g^2r^2)(1-g^2y^2)}{1-a^2g^2}dt^2+\fft{r^2+y^2}{\Delta_r}dr^2+\fft{r^2+y^2}{1-g^2y^2}d\theta_1^2 \right. \\& \left.
		\quad+r^2\sin^2\theta_1(d\theta_2^2+\sin^2\theta_2d\phi_1^2+\cos^2\theta_2d\phi_2^2)+\fft{r^2+a^2}{1-a^2g^2}\cos^2\theta_1d\phi_3^2 \right. \\
		&\left. \quad+\fft{2m's'_1{}^2 k_1^2}{r^2(r^2+y^2)H_1(1-a^2g^2)^2(s'_1{}^2-s'_2{}^2)}+\fft{2m's'_2{}^2 k_2^2}{r^2(r^2+y^2)H_2(1-a^2g^2)^2(s'_2{}^2-s'_1{}^2)}\, \right).
	\end{split}\label{sol2:metric}
\end{equation}
The 1-forms read\footnote{In (3.16) of \cite{Chow:2011fh} the factors of $\tc_I$'s are missing, see (6-7) of \cite{Wu:2011gp} for the correct expressions.}
\begin{equation}
	\begin{split}
		A^1_{(1)}&=\fft{2m's'_1}{H_1(r^2+y^2)r^2(1-a^2g^2)}k_1\,,\\
		A^2_{(1)}&=\fft{2m's'_2}{H_2(r^2+y^2)r^2(1-a^2g^2)}k_2\,.
	\end{split}\label{sol2:1form}
\end{equation}
The 3-form potential is
\begin{equation}
	A_{(3)}=-\fft{2m's'_1s'_2a}{r^2+y^2}\sin^4\theta_1\,\text{vol}_{S^3}-\fft{2m's'_1s'_2ag}{(1-a^2g^2)r^2}\sin\theta_1\cos\theta_1 dt\wedge d\theta_1\wedge d\phi_3\,,\label{sol2:3form}
\end{equation}
where the 3-sphere volume form (\ref{sol2:metric}) can be written explicitly as
\begin{equation}
	\text{vol}_{S^3}=\sin\theta_2\cos\theta_2 d\theta_2\wedge d\phi_1\wedge d\phi_2\,.
\end{equation}
Finally, the 2-form potential is\footnote{In (3.21) of \cite{Chow:2011fh} the factor of $r^2+a^2$ in the denominator should be replaced with $r^2$ as in (8) of \cite{Wu:2011gp}.} 
\begin{equation}
	A_{(2)}=\left(\fft{1}{H_1}+\fft{1}{H_2}\right)\fft{m's'_1s'_2(1-g^2y^2)a}{(r^2+y^2)r^2(1-a^2g^2)}\cos^2\theta_1 dt\wedge d\phi_3\,.\label{sol2:2form}
\end{equation}
In the above solution we have used the expressions
\begin{equation}
	\begin{split}
		s'_I&=\sinh\delta'_I\,,\quad c'_I=\cosh\delta'_I\,,\\
		\tc'_I&=\sqrt{1+a^2g^2s'_I{}^2}\,,\\
		H_I(r,y)&=1+\fft{2m's'_I{}^2}{(r^2+y^2)r^2}\,,\\
		\Delta_r(r)&=\left(r^2+a^2-\fft{2m'}{r^2}\right)\left(1+g^2r^2-\fft{2m'g^2s'_1{}^2s'_2{}^2}{r^2}\right)+2m'g^2c'_1{}^2c'_2{}^2\,,\\
		\Delta_y(y)&=(1-g^2y^2)(a^2-y^2)\,,\\
		\mV_I(r)&=1+g^2r^2+\fft{2m's'_I{}^2g^2}{r^2}\,,\\
		V_y(y)&=1-g^2y^2\,,\\
		\mW_I(r)&=1+\fft{r^2}{a^2}+\fft{2m's'_I{}^2}{r^2a^2}\,,\\
		\tV_y(y)&=1-\fft{y^2}{a^2}\,,\\
		k_1&=c'_1\tc'_2(1-g^2y^2)dt-c'_2\tc'_1a\cos^2\theta_1d\phi_3\,,\\
		k_2&=c'_2\tc'_1(1-g^2y^2)dt-c'_1\tc'_2a\cos^2\theta_1d\phi_3\,.
	\end{split}\label{sol2:functions:prime}
\end{equation}
Note that here we have been using primed parameters distinguished from the unprimed parameters used in the solution presented in Section~\ref{sec:BHsol}.

\medskip
\noindent\textbf{Comparison}
\medskip

\noindent 
For a comparison with the solution in Section~\ref{sec:BHsol}, we first implement the change of parameters
\begin{equation}
	\fft{c_1-\fft12a^2g^2(c_1-c_2)}{s_1}=\fft{c'_1\tc'_2}{s'_1}\,,\quad \fft{c_2+\fft12a^2g^2(c_1-c_2)}{s_2}=\fft{c'_2\tc'_1}{s'_2}\,,\quad ms_I^2=m's'_I{}^2\,.\label{prime:unprime}
\end{equation}
Under the reparametrization (\ref{prime:unprime}), the 1-rotation \& 2-charges solution \eqref{sol2:metric}, \eqref{sol2:1form}, \eqref{sol2:3form} and \eqref{sol2:2form} can be written as
\begin{equation}
	\begin{split}
		ds^2&=(H_1H_2)^{\fft15}\left(-\fft{(1+g^2r^2)(1-g^2y^2)}{1-a^2g^2}dt^2+\fft{r^2+y^2}{\Delta_r}dr^2+\fft{r^2+y^2}{1-g^2y^2}d\theta_1^2 \right. \\
		&\left. \quad+r^2\sin^2\theta_1(d\theta_2^2+\sin^2\theta_2d\phi_1^2+\cos^2\theta_2d\phi_2^2)+\fft{r^2+a^2}{1-a^2g^2}\cos^2\theta_1d\phi_3^2 \right. \\
		&\left.\quad+\fft{1-\fft{1}{H_1}}{1-(s_2/s_1)^2}K_1^2+\fft{1-\fft{1}{H_2}}{1-(s_1/s_2)^2}K_2^2\right)\,,\\
		A^I_{(1)}&=\bigg(1-\fft{1}{H_I}\bigg)K_I\,,\\
		A_{(2)}&=\left(\fft{1}{H_1}+\fft{1}{H_2}\right)\fft{ms_1s_2(1-g^2y^2)a}{(r^2+y^2)r^2(1-a^2g^2)}\cos^2\theta_1 dt\wedge d\phi_3\,,\\
		A_{(3)}&=-\fft{2ms_1s_2a}{r^2+y^2}\sin^4\theta_1\,\text{vol}_{S^3}-\fft{2ms_1s_2ag}{(1-a^2g^2)r^2}\sin\theta_1\cos\theta_1 dt\wedge d\theta_1\wedge d\phi_3\,,
	\end{split}\label{sol2}
\end{equation}
where we have introduced
\begin{equation}
	\begin{split}
		K_1&=\fft{(c_1-\fft12a^2g^2(c_1-c_2))(1-g^2y^2)dt-(c_2+\fft12a^2g^2(c_1-c_2))a\cos^2\theta_1d\phi_3}{s_1(1-a^2g^2)}\,,\\
		K_2&=\fft{(c_2+\fft12a^2g^2(c_1-c_2))(1-g^2y^2)dt-(c_1-\fft12a^2g^2(c_1-c_2))a\cos^2\theta_1d\phi_3}{s_2(1-a^2g^2)}\,.
	\end{split}
\end{equation}
Now it is straightforward to check that the solution in Section~\ref{sec:BHsol} reduces to the above 1-rotation \& 2-charges solution (\ref{sol2}) in the limit
\begin{equation}
	\lim_{a_2\to0}\lim_{a_1\to0}\lim_{a_3\to a}\,.\label{sol2:reduction}
\end{equation}
The thermodynamic quantities of solution evaluated in Section~\ref{sec:analysis:THD} reduce to the ones of the 1-rotation \& 2-charges solution given in \cite{Chow:2011fh,Wu:2011gp} as\footnote{There is a factor of $\fft12$ typo in the temperature given in (4.5) of \cite{Chow:2011fh}.}
\begin{equation}
	\begin{alignedat}{3}
		T&=T^\text{\cite{Wu:2011gp}}=\fft12T^\text{\cite{Chow:2011fh}}\,,&\quad\Omega_i&=(0,0,\Omega)^\text{\cite{Chow:2011fh,Wu:2011gp}}\,,&\quad \Phi_I&=\Phi_I^\text{\cite{Chow:2011fh,Wu:2011gp}}\,,\\
		S&=\fft{1}{G_N}S^\text{\cite{Chow:2011fh,Wu:2011gp}}\,,&\quad J_i&=\fft{1}{G_N}(0,0,J)^\text{\cite{Chow:2011fh,Wu:2011gp}}\,,&\quad Q_I&=\fft{1}{G_N}Q_I^\text{\cite{Chow:2011fh,Wu:2011gp}}\,,\\
		E&=\fft{1}{G_N}E^\text{\cite{Chow:2011fh,Wu:2011gp}}\,,
	\end{alignedat}\label{general:to:case2}
\end{equation}
under the limit (\ref{sol2:reduction}) and the reparametrization of constants (\ref{prime:unprime}) where we have restored the Newton constant explicitly. Note that the sign flip $g\to-g$ should be taken on the RHS of \eqref{general:to:case2} considering the convention difference (\ref{case2:convention}) but the thermodynamic quantities are in fact invariant under $g\to-g$ in the 1-rotation limit (\ref{sol2:reduction}).

\subsection{3-equal-rotations \& 2-charges}\label{App:limits:case3}
Finally we review the 3-equal-rotations \& 2-charges black hole solution given in \cite{Chong:2004dy}. Before writing down the solution, we first implement the coordinate transformation
\begin{equation}
	\text{i})~t\to(1-ag)t\,,\qquad 	\text{ii})~\tau\to\tau -gt\,, \label{Chong04:to:Cvetic05}
\end{equation}
following \cite{Cvetic:2005zi}. Since the supergravity conventions of \cite{Chow:2007ts} summarized in Section~\ref{sec:7dsugra} are different from those of \cite{Chong:2004dy,Cvetic:2005zi,Choi:2018hmj} as
\begin{equation}
	\text{\cite{Chow:2007ts}}=\text{\cite{Chong:2004dy,Cvetic:2005zi,Choi:2018hmj}}\big|_{g\to-g}\,,\label{case3:convention}
\end{equation}
we further flip the sign of the gauge coupling after implementing the coordinate transformation \eqref{Chong04:to:Cvetic05} as
\begin{equation}
	\text{iii})~g\to-g\,.
\end{equation}
We also rename the coordinates of \cite{Chong:2004dy} as
\begin{equation}
	(t,r,\psi,\xi,\varphi_1,\varphi_2,\varphi_3)=(t,\sqrt{\rho^2-a^2},\tau,\xi,\varphi_1,\varphi_2,\varphi_3)^\text{\cite{Chong:2004dy}}\,,\label{sol3:coords}
\end{equation}
where we have introduced the $\varphi_i$ angles for the left-invariant SU(2) 1-forms
\begin{equation}
	\begin{split}
		\sigma_1&=\cos\varphi_3 d\varphi_1+\sin\varphi_1\sin\varphi_3 d\varphi_2\,,\\
		\sigma_2&=\sin\varphi_3 d\varphi_1-\sin\varphi_1\cos\varphi_3 d\varphi_2\,,\\
		\sigma_3&=d\varphi_3+\cos\varphi_1 d\varphi_2\,,
	\end{split}
\end{equation}
satisfying $d\sigma_i=-\fft12\epsilon_{ijk}\sigma_j\wedge\sigma_k$. Several typos in \cite{Chong:2004dy,Cvetic:2005zi} were corrected in \cite{Choi:2018hmj} and we take these corrections into account below. 

The metric reads
\begin{equation}
	\begin{split}
		ds^2	&=(H_1H_2)^{\fft15}\left(-\fft{W}{f_1}dt^2+\fft{r^2(r^2+a^2)^2}{W}dr^2+\fft{r^2+a^2}{1-a^2g^2}ds^2_{\mathbb{CP}^2}\right.\\
		&\left.\quad+\fft{f_1}{(r^2+a^2)^2H_1H_2(1-a^2g^2)^2}\left(\sigma+gdt-\fft{2f_2}{f_1}(1+ag)dt\right)^2\right)\\
		&=(H_1H_2)^{\fft15}\left(-\fft{1+g^2r^2}{1-a^2g^2}dt^2+\fft{r^2(r^2+a^2)^2}{W}dr^2+\fft{r^2+a^2}{1-a^2g^2}\Big(\sigma^2+ds^2_{\mathbb{CP}^2}\Big)\right.\\
		&\left.\quad+\fft{2ms_1^2}{(s_1^2-s_2^2)(r^2+a^2)^2H_1(1-a^2g^2)^2}\Big(\alpha_1(1+ag)dt-a\alpha_2(\sigma+gdt)\Big)^2\right.\\
		&\left.\quad+\fft{2ms_2^2}{(s_2^2-s_1^2)(r^2+a^2)^2H_2(1-a^2g^2)^2}\Big(\alpha_2(1+ag)dt-a\alpha_1(\sigma+gdt)\Big)^2\right)\,,
	\end{split}\label{sol3:metric}
\end{equation}
where the 2nd expression is from Appendix B of \cite{Wu:2011gp} with the convention differences \eqref{case2:convention} and \eqref{case3:convention} taken into account. The $\mathbb{CP}_2$ metric and the U(1) fibration specified by the 1-form $\sigma$ over $\mathbb{CP}_2$ in the metric (\ref{sol3:metric}) can be written explicitly as
\begin{equation}
	\begin{split}
		\sigma&=d\psi+\fft12\sin^2\xi\,\sigma_3\,,\\
		ds^2_{\mathbb{CP}^2}&=d\xi^2+\fft14\sin^2\xi(\sigma_1^2+\sigma_2^2)+\fft14\sin^2\xi\cos^2\xi \sigma_3^2\,,
	\end{split}\label{sigma:CP2}
\end{equation}
in the Fubini-Study coordinates. The 1-forms read
\begin{equation}
	\begin{split}
		A^1_{(1)}&=\fft{2ms_1}{(1-a^2g^2)H_1(r^2+a^2)^2}(\alpha_1(1+ag)dt-a\alpha_2(\sigma+gdt))\,,\\
		A^2_{(1)}&=\fft{2ms_2}{(1-a^2g^2)H_2(r^2+a^2)^2}(\alpha_2(1+ag)dt-a\alpha_1(\sigma+gdt))\,.\label{sol3:1form}
	\end{split}
\end{equation}
The 3-form potential and 2-form potential are given by
\begin{align}
	A_{(3)}&=\fft{mas_1s_2}{(1-a^2g^2)(1+ag)(r^2+a^2)}(\sigma+gdt)\wedge d\sigma\,,\label{sol3:3form}
	\\
	A_{(2)}&=\left(\fft{1}{H_1}+\fft{1}{H_2}\right)\fft{mas_1s_2}{(1+ag)(r^2+a^2)^2}dt\wedge\sigma\,.\label{sol3:2form}
\end{align}
In the above solution we have used the expressions
\begin{equation}
	\begin{split}
		s_I&=\sinh\delta_I\,,\quad c_I=\cosh\delta_I\,,\\
		\alpha_1&=c_1-\fft12(1-(1-ag)^2)(c_1-c_2)\,,\\
		\alpha_2&=c_2+\fft12(1-(1-ag)^2)(c_1-c_2)\,,\\
		H_I(r)&=1+\fft{2ms_I^2}{(r^2+a^2)^2}\,,\\
		f_1(r)&=(1-a^2g^2)H_1H_2(r^2+a^2)^3-\fft{4(1-ag)^2m^2a^2s_1^2s_2^2}{(r^2+a^2)^2}\\
		&\quad+\fft12ma^2\left(4(1-ag)^2+2(1-(1-ag)^4)c_1c_2+(1-(1-ag)^2)^2(c_1^2+c_2^2)\right)\,,\\
		f_2(r)&=\fft12g(1-ag)H_1H_2(r^2+a^2)^3 \\&\quad +\fft14ma\left(2(1+(1-ag)^4)c_1c_2+(1-(1-ag)^4)(c_1^2+c_2^2)\right)\,,\\
		W(r)&=g^2H_1H_2(r^2+a^2)^4+(1-a^2g^2)(r^2+a^2)^3+\fft12ma^2\left(4(1-ag)^2 \right. \\& \left. \quad + 2(1-(1-ag)^4)c_1c_2 +(1-(1-ag)^2)^2(c_1^2+c_2^2)\right)\\
		&\quad-\fft12m(r^2+a^2)\left(4(1-a^2g^2)+2a^2g^2(6-8ag+3a^2g^2)c_1c_2 \right. \\& \quad \left. -a^2g^2(2-ag)(2-3ag)(c_1^2+c_2^2)\right)\,.
	\end{split}
\end{equation}
%
\medskip
\noindent\textbf{Comparison}
\medskip

\noindent 
For a comparison with the solution in Section~\ref{sec:BHsol}, we first implement the coordinate transformation
\begin{equation}
	(\psi,\xi,\varphi_1,\varphi_2,\varphi_3)=(-\phi_3,\theta_1,2\theta_2,\phi_1-\phi_2,2\phi_3-\phi_1-\phi_2)\,,\label{FS:to:typical}
\end{equation}
upon which (\ref{sigma:CP2}) reads
\begin{equation}
	\begin{split}
		\sigma&=-\cos^2\theta_1 d\phi_3-\sin^2\theta_1(\sin^2\theta_2 d\phi_1+\cos^2\theta_2 d\phi_2)\,,\\
		ds^2_{\mathbb{CP}^2}&=d\theta_1^2+\sin^2\theta_1d\theta_2^2+\sin^2\theta_1\sin^2\theta_2\cos^2\theta_2(d\phi_1-d\phi_2)^2\\
		&\quad+\sin^2\theta_1\cos^2\theta_1(d\phi_3-\sin^2\theta_2d\phi_1-\cos^2\theta_2d\phi_2)^2\,.
	\end{split}
\end{equation}
Under the coordinate transformation \eqref{FS:to:typical}, the 3-equal-rotations and 2-charges solution \eqref{sol3:metric}, \eqref{sol3:1form}, \eqref{sol3:3form} and \eqref{sol3:2form} is rewritten as
\begin{equation}
	\begin{split}
		ds^2&=(H_1H_2)^{\fft15}\left(-\fft{1+g^2r^2}{1-a^2g^2}dt^2+\fft{r^2(r^2+a^2)^2}{W}dr^2+\fft{1-\fft{1}{H_1}}{1-(s_2/s_1)^2}K_1^2+\fft{1-\fft{1}{H_2}}{1-(s_1/s_2)^2}K_2^2 \right. \\
		&\left. +\fft{r^2+a^2}{1-a^2g^2}\Big(d\theta_1^2+\sin^2\theta_1d\theta_2^2+\sin^2\theta_1\sin^2\theta_2d\phi_1^2+\sin^2\theta_1\cos^2\theta_2d\phi_2^2+\cos^2\theta_1d\phi_3^2\Big) \right)\,,\\
		A^I_{(1)}&=\bigg(1-\fft{1}{H_I}\bigg)K_I\,,\\
		A_{(3)}&=\fft{mas_1s_2}{(1-a^2g^2)(1+ag)(r^2+a^2)}(\sigma+gdt)\wedge d\sigma\,,\\
		A_{(2)}&=\left(\fft{1}{H_1}+\fft{1}{H_2}\right)\fft{mas_1s_2}{(1+ag)(r^2+a^2)^2}dt\wedge\sigma\,,
	\end{split}\label{sol3}
\end{equation}
where we have defined
\begin{equation}
	\begin{split}
		K_1&=\fft{1}{s_1(1-a^2g^2)}\Big(\alpha_1(1+ag)dt-a\alpha_2(\sigma+gdt)\Big)\,,\\
		K_2&=\fft{1}{s_2(1-a^2g^2)}\Big(\alpha_2(1+ag)dt-a\alpha_1(\sigma+gdt)\Big)\,.
	\end{split}
\end{equation}
Now it is straightforward to check that the solution in Section~\ref{sec:BHsol} reduces to the 3-equal-rotations \& 2-charges solution (\ref{sol3}) in the limit
\begin{equation}
	\lim_{a_3\to-a}\lim_{a_2\to-a}\lim_{a_1\to-a}\,.\label{sol3:reduction}
\end{equation}
The thermodynamic quantities of the solution evaluated in Section~\ref{sec:analysis:THD} reduce to the ones of the 3-equal-rotations \& 2-charges black hole solution given in \cite{Cvetic:2005zi} whose typos are corrected in \cite{Choi:2018hmj}. In terms of the corrected thermodynamic quantities in \cite{Choi:2018hmj} we find
\begin{equation}
	\begin{alignedat}{3}
		T&=[gT^\text{\cite{Choi:2018hmj}}]_{g\to-g}\,,&\qquad\Omega_i&=[g\Omega^\text{\cite{Choi:2018hmj}}]_{g\to-g}\,,&\qquad \Phi_I&=[\fft12\Phi^\text{\cite{Choi:2018hmj}}]_{g\to-g}\,,\\
		S&=[S^\text{\cite{Choi:2018hmj}}]_{g\to-g}\,,&\qquad J_i&=[J^\text{\cite{Choi:2018hmj}}]_{g\to-g}\,,&\qquad Q_I&=[2gQ_I^\text{\cite{Choi:2018hmj}}]_{g\to-g}\,,\\
		E&=[gE^\text{\cite{Choi:2018hmj}}]_{g\to-g}\,,
	\end{alignedat}\label{general:to:case3}
\end{equation}
in the limit \eqref{sol3:reduction}. Consequently the modified supergravity angular velocities and electrostatic potentials \eqref{eq:varphiomega} are related to those of \cite{Choi:2018hmj} given in their (4.39) as
\begin{equation}
	\begin{split}
		\omega_i&=\fft{\Omega_i+g}{T}=\fft{[g\Omega^\text{\cite{Choi:2018hmj}}]_{g\to-g}+g}{[gT^\text{\cite{Choi:2018hmj}}]_{g\to-g}}=\fft{\Omega^\text{\cite{Choi:2018hmj}}-1}{T^\text{\cite{Choi:2018hmj}}}\bigg|_{g\to-g}=-\zeta^\text{\cite{Choi:2018hmj}}|_{g\to-g}\,,\\ \varphi_I&=\fft{\Phi_I-1}{T}=\fft{[\fft12\Phi_I^\text{\cite{Choi:2018hmj}}]_{g\to-g}-1}{[gT^\text{\cite{Choi:2018hmj}}]_{g\to-g}}=\fft{\Phi_I^\text{\cite{Choi:2018hmj}}-2}{2gT^\text{\cite{Choi:2018hmj}}}\bigg|_{g\to-g}=-\fft{1}{2g}\xi_I^\text{\cite{Choi:2018hmj}}\bigg|_{g\to-g}\,,
		\label{general:to:case3:susy}
	\end{split}
\end{equation}
in the limit (\ref{sol3:reduction}). Note that the sign flip $g\to-g$ is taken on the RHS of \eqref{general:to:case3} and \eqref{general:to:case3:susy} considering the convention difference \eqref{case3:convention}.

\section{An equivalent expression for the black hole metric}\label{App:equivalent}
The black hole metric \eqref{ansatz:metric} can be rewritten as
\begin{equation}
	\begin{split}
		ds^2&=ds_5^2+(H_1(r,y,z)H_2(r,y,z))^{\fft15}\fft{(r^2+y^2)(y^2-z^2)y^2}{(1-g^2y^2)(a_1^2-y^2)(a_2^2-y^2)(a_3^2-y^2)}dy^2\\
		&\quad+(H_1(r,y,z)H_2(r,y,z))^{\fft15}\fft{(r^2+z^2)(z^2-y^2)z^2}{(1-g^2z^2)(a_1^2-z^2)(a_2^2-z^2)(a_3^2-z^2)}dz^2\,,\\
		ds_5^2&=-g_{00}(r,y,z)dt^2+\sum_{i,j=1}^3g_{ij}(r,y,z)(d\phi_i-W_i(r,y,z)dt)(d\phi_j-W_j(r,y,z)dt)\\
		&\quad+(H_1(r,y,z)H_2(r,y,z))^{\fft15}\fft{(r^2+y^2)(r^2+z^2)}{U}dr^2\,,
	\end{split}\label{ansatz:metric:2}
\end{equation}
where $g_{00},g_{ii},g_{ij}=g_{ji}$ and $W_i$ are complicated functions of the coordinates $\{r,y,z\}$ and the parameters $\{m,a_1,a_2,a_3,\delta_1,\delta_2\}$. However, it is still illuminating to present $g_{00}$ and $W_i$ explicitly as ($i,j,k,\in\{1,2,3\}$ are all different)
\begin{align}
	\label{eq:g00}
	g_{00}(r,y,z)&=\fft{(H_1(r,y,z)H_2(r,y,z))^{\fft15}r^4(r^2+y^2)(r^2+z^2)U(r)}{\mS(r)}\left[1+\tfrac{Q(r,y,z)}{P(r,y,z)}U(r)\right],
	\\
	\label{eq:W}
	W_i(r,y,z)&=\fft{\fft{1+g^2r^2}{r^2}(1-g^2y^2)(1-g^2z^2)\mS(r)}{P(r,y,z)}\left[w_i(r)-a_ir^2(r^2+a_j^2)(r^2+a_k^2)\tfrac{U(r)}{\mS(r)}\right],
\end{align}
%
%
in terms of $w_i(r)$ given in (\ref{eq:Omega}) and the newly defined functions
\begin{equation}
	\begin{split}
		Q(r,y,z)&=(1+g^2r^2)\bigg[a_1^2a_2^2a_3^2(1-g^2(r^2+y^2+z^2)+g^4(r^2+y^2)(r^2+z^2))\\
		&\quad+(a_1^2a_2^2+a_2^2a_3^2+a_3^2a_1^2)r^2(1-g^2(r^2+y^2+z^2))\\
		&\quad+(a_1^2+a_2^2+a_3^2)r^2(r^2+g^2y^2z^2)+r^2(r^2(r^2+g^2y^2z^2)-(r^2+y^2)(r^2+z^2))\bigg]\,,\\
		P(r,y,z)&=\fft{1+g^2r^2}{r^2}(1-g^2y^2)(1-g^2z^2)\mS(r)\\
		&\quad+\bigg[r^2(r^2+y^2)(r^2+z^2)\Xi_1\Xi_2\Xi_3-(1-g^2y^2)(1-g^2z^2)\prod_{i=1}^3(r^2+a_i^2)\bigg]U(r)\,.
	\end{split}
\end{equation}
Note that \eqref{eq:g00} and \eqref{eq:W} can be expanded as (recall $\Omega_i=w_i(r_+)$)
\begin{subequations}
\begin{align}
	g_{00}(r,y,z)&=\fft{(H_1(r_+,y,z)H_2(r_+,y,z))^{\fft15}r_+^4(r_+^2+y^2)(r_+^2+z^2)}{\mS(r_+)}U'(r_+)\rho^2+\mO(\rho^4)\,,\\
	W_i(r_+,y,z)&=\Omega_i+\mO(\rho^2)\,,
\end{align}
\end{subequations}
in the near horizon limit
\begin{equation}
	r=r_++\rho^2\qquad(\rho^2\ll r_+)\,.\label{near-horizon}
\end{equation}
Hence the 5d part of the metric \eqref{ansatz:metric:2} can be written simply as
\begin{equation}
\begin{split}
	&ds_5^2\big|_{\rho^2\ll r_+}\\&=(H_1(r_+,y,z)H_2(r_+,y,z))^\fft15(r_+^2+y^2)(r_+^2+z^2)\bigg[-\fft{r_+^4}{\mS(r_+)}U'(r_+)\rho^2dt^2+\fft{4d\rho^2}{U'(r_+)}\bigg]\\
	&\quad+\sum_{i,j=1}^3g_{ij}(r_+,y,z)(d\phi_i-\Omega_idt)(d\phi_j-\Omega_jdt)\,,
\end{split}
\end{equation}
in the near horizon limit \eqref{near-horizon}. This presentation of the metric is convenient for studying its near-horizon properties.


\bibliography{AdS7-BH_JHEP}
\bibliographystyle{JHEP}

\end{document}